\documentclass[a4paper,oneside,11pt]{article}
\pdfoutput=1
\usepackage{cprform}
\usepackage{amsmath,amstext,amsfonts,amsbsy,amssymb,amscd,slashed}
\usepackage{pdflscape,color,svg,epsfig,cite,hyperref,bbm,multirow}
\usepackage{graphicx}
\usepackage{tabularx}
\usepackage{booktabs}
\usepackage{floatpag}

\newcommand{\ba}{\begin{array}}
\newcommand{\ea}{\end{array}}

\newcommand{\req}[1]{Eq.~(\ref{#1})}
\newcommand{\res}[1]{Section~\ref{#1}}

\newcommand{\ret}[1]{Table~\ref{#1}}

\newcommand{\dif}{{\rm d}}

\newcommand{\Dslash}{\relax{\kern+.25em / \kern-.70em D}}

\newcommand{\MeV}{{\rm MeV}}
\newcommand{\GeV}{{\rm GeV}}

\newcommand{\Real}{\relax{\mathsf{\Gamma\kern-.35em R}}}
\newcommand{\Int}{\relax{\mathsf{Z\kern-.40em Z}}}

\newcommand{\half}{{\scriptstyle{{1\over 2}}}}

\newcommand{\NF}{N_\mathrm{\scriptstyle f}}

\newcommand{\gbar}{\kern1pt\overline{\kern-1pt g\kern-0pt}\kern1pt}
\newcommand{\mbar}{\kern2pt\overline{\kern-1pt m\kern-1pt}\kern1pt}
\newcommand{\obar}[1]{\kern3pt\overline{\kern-2pt #1\kern-0pt}\kern1pt}

\newcommand{\oren}[1]{#1_{\rm R}}
\newcommand{\lQCD}{\Lambda_{\rm\scriptscriptstyle QCD}}
\newcommand{\mrgi}{M}

\newcommand{\hopc}{\kappa_{\rm c}}

\newcommand{\fP}{f_{\rm\scriptscriptstyle P}}

\newcommand{\fA}{f_{\rm\scriptscriptstyle A}}

\newcommand{\Zm}{Z_{\rm m}}
\newcommand{\ZP}{Z_{\rm\scriptscriptstyle P}}

\newcommand{\ZA}{Z_{\rm\scriptscriptstyle A}}

\newcommand{\sigmaP}{\sigma_{\rm\scriptscriptstyle P}}

\newcommand{\SigmaP}{\Sigma_{\rm\scriptscriptstyle P}}

\newcommand{\SigmaPI}{\SigmaP^{\rm\scriptscriptstyle I}}
\newcommand{\deltaP}{\delta_{\rm\scriptscriptstyle P}}

\newcommand{\Oa}{\mbox{O}(a)}
\newcommand{\Oasq}{\mbox{O}(a^2)}

\newcommand{\icsw}{c_{\rm sw}}
\newcommand{\ict}{c_{\rm t}}
\newcommand{\icttil}{\tilde c_{\rm t}}

\newcommand{\icA}{c_{\rm\scriptscriptstyle A}}

\newcommand{\abar}{\kern1pt\overline{\kern-1pt a\kern-0.5pt}\kern1pt}

\newcommand{\cO}{{\cal O}}
\newcommand{\cP}{{\cal P}}

\newcommand{\cZ}{{\cal Z}}

\newcommand{\vx}{\mathbf{x}}
\newcommand{\vy}{\mathbf{y}}
\newcommand{\vz}{\mathbf{z}}

\newcommand{\vk}{\mathbf{k}}

\newcommand{\HEs}{{\rm SF}}
\newcommand{\LEs}{{\rm GF}}
\newcommand{\mupt}{\mu_{\rm\scriptscriptstyle pt}}
\newcommand{\muhad}{\mu_{\rm\scriptscriptstyle had}}
\newcommand{\muswi}{\mu_0}
\newcommand{\uSF}{u_{\rm\scriptscriptstyle SF}}
\newcommand{\uGF}{u_{\rm\scriptscriptstyle GF}}
\newcommand{\uhad}{u_{\rm\scriptscriptstyle had}}
\newcommand{\ZM}{Z_{\rm\scriptscriptstyle M}}
\newcolumntype{C}{>{$}c<{$}}
\newcolumntype{R}{>{$}r<{$}}
\newcolumntype{L}{>{$}l<{$}}

\def\Nms{N_{\rm ms}}
\def\tms{\tau_{\rm ms}}
\newcommand{\tint}[1][{}]{\tau_{\rm int}[#1]}

\begin{document}


\begin{titlepage}


\vspace*{-30truemm}
\begin{flushright}
CERN-TH-2018-029\\
IFT-UAM/CSIC-18-011\\
FTUAM-18-4
\end{flushright}
\vspace{15truemm}


\centerline{\Large Non-perturbative quark mass renormalisation and running in $\NF=3$ QCD }
\vskip 10 true mm
\begin{center}
\includesvg[svgpath=figs/,width=0.16\textwidth]{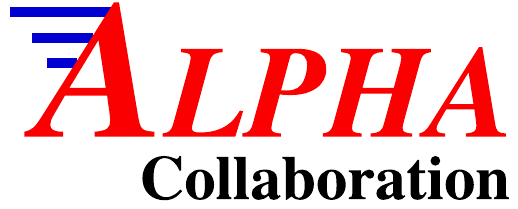}\\
\end{center}

\vskip -2 true mm
\centerline{\bigrm  I.~Campos$^a$, P.~Fritzsch$^b$, C.~Pena$^c$, D.~Preti$^d$, A.~Ramos$^e$, A.~Vladikas$^f$}
\vskip 4 true mm
\centerline{\it $^a$ Instituto de F\'{\i}sica de Cantabria IFCA-CSIC}
\centerline{\it Avda. de los Castros s/n, E-39005 Santander, Spain}
\vskip 3 true mm
\centerline{\it $^b$ Theoretical Physics Department, CERN}
\centerline{\it CH-1211 Geneva 23, Switzerland}
\vskip 3 true mm
\centerline{\it $^c$ Instituto de F\'{\i}sica Te\'orica UAM-CSIC \&} 
\centerline{\it Dpto. de F\'{\i}sica Te\'orica, Universidad Aut\'onoma de Madrid}
\centerline{\it Cantoblanco E-28049 Madrid, Spain}
\vskip 3 true mm
\centerline{\it $^d$ INFN, Sezione di Torino}
\centerline{\it Via Pietro Giuria 1, I-10125 Turin, Italy}
\vskip 3 true mm
\centerline{\it $^e$ School of Mathematics, Trinity College Dublin}
\centerline{\it Dublin 2, Ireland}
\vskip 3 true mm
\centerline{\it $^f$ INFN, Sezione di Tor Vergata}
\centerline{\it c/o Dipartimento di Fisica, Universit\`a di Roma ``Tor Vergata''}
\centerline{\it Via della Ricerca Scientifica 1, I-00133 Rome, Italy}
\vskip 15 true mm


\noindent{\tenbf Abstract:}
We determine from first principles the quark mass anomalous dimension in $\NF=3$ QCD
between the electroweak and hadronic scales. This allows for a fully non-perturbative
connection of the perturbative and non-perturbative regimes of the Standard Model
in the hadronic sector.
The computation is carried out to high accuracy, employing massless $\Oa$-improved
Wilson quarks and finite-size scaling techniques.
We also provide the matching factors required in the renormalisation of
light quark masses from lattice computations with $\Oa$-improved Wilson fermions
and a tree-level Symanzik improved gauge action.
The total uncertainty due to renormalisation and running in the determination
of light quark masses in the SM is thus reduced to about $1\%$.

\vspace{10truemm}

\eject
\end{titlepage}

\tableofcontents

\cleardoublepage

\section{Introduction}
\label{sec:int}

In the paradigm provided by the Standard Model (SM) of Particle Physics,
quark masses are fundamental constants of Nature. More specifically,
Quantum Chromodynamics (QCD), the part of the SM that describes the fundamental
strong interaction, is uniquely defined by the values of the quark masses and the
strong coupling constant. Apart from this intrinsic interest, precise knowledge
of the values of quark masses is crucial for the advancement of frontier research
in particle physics --- one good illustration being the fact that the values
of the bottom and charm quark masses are major sources of uncertainty
in several important Higgs branching fractions, e.g.,
$\Gamma(H\to b\bar b)$ and $\Gamma(H\to c\bar c)$~\cite{Denner:2011mq,Heinemeyer:2013tqa,Almeida:2013jfa,Lepage:2014fla,Petrov:2015jea}.

Quark masses are couplings in the QCD Lagrangian, and have to be treated within
a consistent definition of the renormalised theory. A meaningful determination
can only be achieved by computing physical observables as a function of quark
masses, and matching the result to the experimental values. A non-perturbative
treatment of QCD is mandatory to avoid the presence of unquantified systematic
uncertainties in such a computation: the asymptotic nature of the perturbative
series, and the strongly coupled nature of the interaction at typical hadronic energy scales, implies
the presence of an irreducible uncertainty in any determination that does not
treat long-distance strong interaction effects from first principles. Lattice QCD (LQCD)
is therefore the best-suited framework for a high-precision determination of quark masses.
Indeed, following the onset of the precision era in LQCD, the uncertainties
on the values of both light and heavy quark masses have dramatically decreased
in recent years~\cite{Aoki:2013ldr,Aoki:2016frl,Bazavov:2009fk,McNeile:2010ji,Bazavov:2010yq,Durr:2010vn,Durr:2010aw,Blossier:2010cr,Fritzsch:2012wq,Carrasco:2013zta,Bernardoni:2013xba,Carrasco:2014cwa,Alexandrou:2014sha,Colquhoun:2014ica,Chakraborty:2014aca,Yang:2014sea,Blum:2014tka}.

The natural observables employed in a LQCD computation of quark masses are
hadronic quantities, considered at energy scales around or below $1~\GeV$.
This requires, in particular, to work out the renormalisation non-perturbatively.
Then, in order to make contact with the electroweak scale, where the masses
are used to compute the QCD contribution to high-energy observables, the masses
have to be run with the Renormalisation Group (RG) across more than two orders
of magnitude in energy. While high-order perturbative estimates of the anomalous
dimension of quark masses in various renormalisation schemes exist~\cite{Chetyrkin:1997dh,Vermaseren:1997fq,Gracey:2003yr},
a non-perturbative determination is mandatory to match the current percent-level
precision of the relevant hadronic observables.

In this work we present a high-precision determination of the anomalous dimension
of quark masses in QCD with three light quark flavours, as well as of the renormalisation
constants required to match bare quark masses.\footnote{Preliminary results have appeared
as conference proceedings in~\cite{Campos:2015fka,Campos:2016vxh,Campos:2016eef}.} This is a companion project of the
recent high-precision determination of the $\beta$ function and the $\lQCD$ parameter
in $\NF=3$ QCD by the ALPHA~Collaboration~\cite{Brida:2016flw,DallaBrida:2018rfy,DallaBrida:2016kgh,Bruno:2017gxd}.
We will employ the Schr\"odinger Functional~\cite{Luscher:1992an,Sint:1993un}
as an intermediate renormalisation scheme that allows to make contact between
the hadronic scheme used in the computation of bare quark masses and the perturbative
schemes used at high energies, and employ well-established finite-size
recursion techniques~\cite{Capitani:1998mq,DellaMorte:2005kg,Guagnelli:2005zc,Palombi:2005zd,Dimopoulos:2006ma,Dimopoulos:2007ht,Palombi:2007dr,Blossier:2010jk,Bernardoni:2014fva,Papinutto:2016xpq,Dimopoulos:2018zef,Pena:2017hct} to compute the RG running non-perturbatively.
Our main result is a high-precision determination of the mass anomalous dimension
between the electroweak scale and hadronic scales at around $200~\MeV$,
where contact with hadronic observables obtained from simulations by the
CLS effort~\cite{Bruno:2014jqa} can be achieved.

The paper is structured as follows. In Section~2 we describe our strategy,
which (similar to the determination of $\lQCD$) involves using two different
definitions of the renormalised coupling at energies above and below an
energy scale around $2~\GeV$. Sections~3 and~4 deal with the determination of the anomalous dimension
above and below that scale, respectively. Section~5 discusses the determination
of the renormalisation constants needed to match to a hadronic scheme at low
energies. Conclusions and outlook come in Section~6. Several technical aspects
of the work are discussed in appendices.

\section{Strategy}
\label{sec:str}

\subsection{Quark running and RGI masses}

QCD is a theory with $\NF+1$ parameters: the coupling constant $g$ and the $\NF$
quark masses $\{m_i,\,i=1,\ldots,\NF\}$. When the theory is defined using some
regularisation, $g$ and $m_i$ are taken to be the bare parameters appearing in
the Lagrangian. Removing the regularisation requires to define renormalised
parameters $\gbar,\{\mbar_i,\,i=1,\ldots,\NF\}$ at some energy scale $\mu$.
In the following we will assume the use of renormalisation conditions that
are independent of the values of the quark masses, which leads to mass-independent
renormalisation schemes. The renormalised parameters are then functions of
the renormalisation scale $\mu$ alone~\cite{Weinberg:1951ss,tHooft:1973mfk},
and their scale evolution is given by renormalisation group equations of the form
\begin{align}
\label{eq:beta}
\mu\frac{\dif}{\dif\mu}\,\gbar(\mu) & = \beta(\gbar(\mu))\,,\\
\label{eq:tau}
\mu\frac{\dif}{\dif\mu}\,\mbar_i(\mu) & = \tau(\gbar(\mu))\,\mbar_i(\mu)\,, \qquad i=1,\ldots,\NF\,.
\end{align}
The renormalisation group functions $\beta$ and $\tau$ admit perturbative expansions
of the form
\begin{align}
\beta(g) & \underset{g \to 0}{\sim} -g^3(b_0+b_1g^2+b_2g^4+\ldots)\,,\\
\tau(g)  & \underset{g \to 0}{\sim} -g^2(d_0+d_1g^2+d_2g^4+\ldots)\,,
\end{align}
with universal coefficients given by~\cite{vanyaterent,Khriplovich:1969aa,thooft,Gross:1973id,Politzer:1973fx,Caswell:1974gg,Jones:1974mm}
\begin{align}
\label{eq:univ}
b_0 & = \frac{1}{(4\pi)^2}\left(11-\frac{2}{3}\NF\right)\,, \quad
b_1   = \frac{1}{(4\pi)^4}\left(102-\frac{38}{3}\NF\right)\,, \\
d_0 & = \frac{8}{(4\pi)^2}\,.
\end{align}
The higher-order coefficients $b_{n \geq 2},~d_{n \geq 1}$ are instead
renormalisation scheme-dependent.
It is trivial to integrate Eqs.~(\ref{eq:beta},\ref{eq:tau}) formally,
to obtain the renormalisation group invariants (RGI)
\begin{align}
\label{eq:lambdadef}
\lQCD & = \mu\left[b_0\gbar^2(\mu)\right]^{-\frac{b_1}{2b_0^2}}\,e^{-\frac{1}{2b_0\gbar^2(\mu)}}\,
\exp\left\{
-\int_0^{\gbar(\mu)}\dif g\left[\frac{1}{\beta(g)}+\frac{1}{b_0g^3}-\frac{b_1}{b_0^2g}\right]
\right\}\,,\\
\label{eq:mrgidef}
\mrgi_i & = \mbar_i(\mu) \left[2b_0\gbar^2(\mu)\right]^{-\frac{d_0}{2b_0}}
\exp\left\{
-\int_0^{\gbar(\mu)}\dif g\left[\frac{\tau(g)}{\beta(g)}-\frac{d_0}{b_0g}\right]
\right\}\,.
\end{align}
Note that the integrands are finite at $g=0$, making the integrals well defined
(and zero at universal order in perturbation theory). Note also that $\lQCD$ and
$\mrgi_i$ are non-perturbatively defined via the previous expressions. 
It is easy to check, furthermore, that they are
$\NF$-dependent but $\mu$-independent. They can be interpreted as the integration constants
of the renormalisation group equations.
Also the ratios $\mbar_i(\mu)/\mbar_j(\mu),\,i\neq j$
are scale-independent. Furthermore, the ratios $\mrgi_i/\mbar_i(\mu)$
are independent of the quark flavour $i$, due to the mass-independence of the
quark mass anomalous dimension $\tau$. Finally, the values of $M_i$ can be easily
checked to be independent of the renormalisation scheme. The value of $\lQCD$
is instead scheme-dependent, but the ratio $\lQCD/\lQCD'$ between its values
in two different schemes can be calculated exactly using one-loop perturbation theory.

\subsection{Step scaling functions}

In our computation, we will access the renormalisation group functions $\beta$
and $\tau$ through the quantities $\sigma$ and $\sigmaP$, defined as
\begin{subequations}
\label{eq:sigmas}
\begin{align}
\label{eq:sigma}
\ln 2 & = -\int_{\sqrt{u}}^{\sqrt{\sigma(u)}}\kern-0pt\dif g\,\frac{1}{\beta(g)}\,,\\
\label{eq:sigmaP}
\sigmaP(u) & = \exp\left\{
-\int_{\sqrt{u}}^{\sqrt{\sigma(u)}}\kern-0pt\dif g\,\frac{\tau(g)}{\beta(g)}
\right\}\,,
\end{align}
\end{subequations}
and to which we will refer as coupling and mass step scaling functions (SSFs), respectively.
They correspond to the renormalisation group evolution operators for the coupling
and quark mass between scales that differ by a factor of two, viz.
\begin{align}
\label{eq:sigmaalt}
\sigma(u) = \left.\gbar^2(\mu/2)\right|_{u=\gbar^2(\mu)}\,, \qquad
\sigmaP(u) = \left.\frac{\mbar_i(\mu)}{\mbar_i(\mu/2)}\right|_{u=\gbar^2(\mu)}\,.
\end{align}
The main advantage of these quantities is that they can be computed
accurately on the lattice, with a well-controlled continuum limit for a very wide range of
energy scales. This is so thanks to the use of finite size scaling
techniques, first introduced for quark masses in~\cite{Capitani:1998mq}. The RG
functions can be non-perturbatively computed between the hadronic
regime and the electroweak scale, establishing safe contact with the
asymptotic perturbative regime

In order to compute the SSF $\sigmaP$, we define renormalised quark masses through
the partially conserved axial current (PCAC) relation,
\begin{gather}
\label{eq:pcac}
\partial_\mu (\oren{A})^{ij}_{\mu} = (\mbar_i + \mbar_j)\oren{P}^{ij}\,,
\end{gather}
where the renormalised, non-singlet ($i \neq j$) axial current and pseudoscalar density
operators are given by
\begin{align}
(\oren{A})^{ij}_{\mu}(x) & = \ZA \bar\psi_i(x)\gamma_\mu\gamma_5\psi_j(x)\,,\\
(\oren{P})^{ij}(x) & = \ZP \bar\psi_i(x)\gamma_5\psi_j(x)\,.
\end{align}
In these expressions $\ZP$ is the renormalisation constant of the pseudoscalar density
in the regularised theory, and $\ZA$ is the finite axial current normalisation, 
required when the QCD regularisation breaks chiral symmetry, as with lattice Wilson fermions.
Eq.~(\ref{eq:pcac}) implies
that, up to the finite current normalisation, current quark masses renormalise with $\ZP^{-1}$.
Therefore, the SSF $\sigmaP$ of Eq.~(\ref{eq:sigmaP}) can be obtained
by computing $\ZP$ at fixed bare gauge coupling $g_0^2$ --- i.e., fixed lattice
spacing --- for scales $\mu$ and $\mu/2$. This is repeated for several different
values of the lattice spacing $a$, and the continuum limit of their ratio is taken, viz.
\begin{gather}
\label{eq:ssfcl}
\sigmaP(u) = \lim_{a \to 0}\left.\SigmaP(g_0^2,a\mu)\right|_{\gbar^2(\mu)=u}\,, \qquad
\SigmaP(g_0^2,a\mu) \equiv \frac{\ZP(g_0^2,a\mu/2)}{\ZP(g_0^2,a\mu)}\,,
\end{gather}
where $g_0^2$ is the bare coupling, univocally related to $a$
in mass-independent schemes. The condition that the value of the renormalised coupling
$u=\gbar^2(\mu)$ is kept fixed in the extrapolation ensures that the latter is taken
along a line of constant physics.\footnote{Note that the assumption of a lattice regularisation
in Eqs.~(\ref{eq:pcac},\ref{eq:ssfcl}) is inessential: the construction can be applied to
any convenient regularisation, provided currents are correctly normalised,
and $\sigmaP$ is obtained by removing the regulator at constant physics.}

In this work we
will determine non-perturbatively $\tau(\bar g)$ from
Eq.~\eqref{eq:sigmaP}. Note that in order to do so, we need the RG
function $\beta(\bar g)$. As we will note later, the non-perturbative
determination of the $\beta$ function has already been done in our
schemes of choice~\cite{Brida:2016flw,DallaBrida:2016kgh}.
In practice, given the $\beta$ function (and hence $\sigma(u)$), and
the lattice results for $\SigmaP(g_0^2,a\mu)$, one determines the
anomalous dimension $\tau(\gbar)$ by extrapolating $\SigmaP(u,a\mu)$
to the continuum (Eq.~\eqref{eq:ssfcl}) and then using the relation
Eq.~\eqref{eq:sigmaP} to constrain the functional form of
$\tau(\gbar)$. 

\subsection{Renormalisation schemes}

In order to control the connection between hadronic observables and RGI quantities,
we will use intermediate finite-volume renormalisation schemes that allow to define fully non-perturbative
renormalised gauge coupling and quark masses. For that purpose, $\ZP$ will be defined
by a renormalisation condition imposed using the Schr\"odinger Functional (SF)~\cite{Luscher:1992an,Sint:1993un}.
In the following, we will
adopt the conventions and notations for the lattice SF setup introduced in~\cite{Luscher:1996sc}.

SF schemes are based on the formulation of QCD in a finite space-time volume of
size $T^3 \times L$, with inhomogeneous Dirichlet boundary conditions
at Euclidean times $x_0=0$ and $x_0=T$. The boundary condition for gauge fields
has the form
\begin{gather}
\left.U_k(x)\right|_{x_0=0}=\cP\exp\left\{
a\int_0^1\dif t \,\,C_k(\vx+(1-t)a\hat\vk)
\right\}\,,
\end{gather}
where $\hat\vk$ is a unit vector in the direction $k$, $\cP\exp$ is a path-ordered
exponential, and $C_k$ is some smooth gauge field. A similar expression applies at
$x_0=T$ in terms of another field $C'_k$.
Fermion fields obey the boundary conditions
\begin{align}
  \left.P_+\psi(x)\right|_{x_0=0\,} &= \rho(\vx) \,, & \left.\bar\psi(x)P_-\right|_{x_0=0\,} &= \bar\rho(\vx) \,, \\
  \left.P_-\psi(x)\right|_{x_0=T} &= \rho'(\vx)\,, & \left.\bar\psi(x)P_+\right|_{x_0=T} &= \bar\rho'(\vx)\,, &
\end{align}
with $P_\pm=\half(\mathbf{1}\pm\gamma_0)$.
Gauge fields are periodic in spatial directions, whereas fermion fields are periodic
up to a global phase
\begin{gather}
\psi(x+L\hat\vk)=e^{i\theta_k}\psi(x)\,, \qquad
\bar\psi(x+L\hat\vk)=\bar\psi(x)e^{-i\theta_k}\,.
\end{gather}
The SF is the generating functional
\begin{gather}
\cZ[C,\bar\rho,\rho;C',\bar\rho',\rho']=\int{\rm D}[U]{\rm D}[\psi]{\rm D}[\bar\psi]\,e^{-S[U,\bar\psi,\psi]}\,,
\end{gather}
where the integral is performed over all fields with the specified boundary values.
Expectation values of any product $\cO$ of fields are then given by
\begin{gather}
\langle\cO\rangle = \left\{
\frac{1}{\cZ}
\int{\rm D}[U]{\rm D}[\psi]{\rm D}[\bar\psi]\,\cO e^{-S[U,\bar\psi,\psi]}
\right\}_{\bar\rho=\rho=\bar\rho'=\rho'=0}\,,
\end{gather}
where $\cO$ can involve, in particular, the ``boundary fields''
\begin{gather}
\zeta(\vx)=\frac{\delta}{\delta\rho(\vx)}\,, \quad
\bar\zeta(\vx)=-\frac{\delta}{\delta\bar\rho(\vx)}\,; \qquad \qquad
\zeta'(\vx)=\frac{\delta}{\delta\rho'(\vx)}\,, \quad
\bar\zeta'(\vx)=-\frac{\delta}{\delta\bar\rho'(\vx)}\,.
\end{gather}
The Dirichlet boundary conditions
provide an infrared cutoff to the possible wavelengths of quark and gluon fields,
which allows to simulate at vanishing quark mass.
The presence of non-trivial boundary conditions requires, in general,
additional counterterms to renormalise the theory~\cite{Symanzik:1981wd,Luscher:1985iu,Luscher:1992an}.
In the case of the SF, it has been shown in~\cite{Sint:1995rb}
that no additional counterterms are needed with respect to the periodic case,
except for one boundary term that amounts to rescaling the boundary values
of quark fields by a logarithmically divergent factor, which is furthermore
absent if $\bar\rho=\rho=\bar\rho'=\rho'=0$. It then follows that the SF
is finite after the usual QCD renormalisation.

The SF naturally allows for the introduction of finite-volume renormalisation
schemes, where the renormalisation scale is identified with the inverse box length,
\begin{gather}
\mu = \frac{1}{L}\,.
\end{gather}
The renormalisation of the pseudoscalar density, and hence of quark masses,
is treated in the same way as in~\cite{Capitani:1998mq}. We introduce the SF correlation functions
\begin{align}
\label{eq:fP}
\fP(x_0) & = -\frac{1}{3}\int\dif^3x \langle P^{ij}(x) \, \cO^{ji}\rangle\,,\\
\label{eq:f1}
f_1 & = -\frac{1}{3}\langle \cO'^{ij}\,\cO^{ji} \rangle\,,
\end{align}
where $P^{ij}$ is the unrenormalised pseudoscalar density, and $\cO$, $\cO'$ are
operators with pseudoscalar quantum numbers made up of boundary quark fields
\begin{gather}
\cO^{ij} = \frac{1}{L^3}\int\dif^3y\int\dif^3z \,\bar\zeta_i(\vy)\gamma_5\zeta_j(\vz)\,, \quad
\cO'^{ij} = \frac{1}{L^3}\int\dif^3y\int\dif^3z \,\bar\zeta'_i(\vy)\gamma_5\zeta'_j(\vz)\,.
\end{gather}
The pseudoscalar renormalisation constant is then defined as
\begin{gather}
\label{eq:ZP}
\ZP\,\frac{\fP(L/2)}{\sqrt{3f_1}} = \left.\frac{\fP(L/2)}{\sqrt{3f_1}}\right|_{\rm tree~level}\,,
\end{gather}
with all correlation functions computed at zero quark masses. The renormalisation
condition is fully specified by fixing the boundary conditions and the box geometry
as follows:
\begin{gather}
T=L\,, \qquad C=C'=0\,, \qquad \theta_k \equiv \theta=\frac{1}{2}\,.
\end{gather}
Furthermore, for computational convenience (cf.~below), all correlation functions
will be computed in a fixed topological sector of the theory, chosen
to be the one with total topological charge $Q=0$. This is just part of the
scheme definition, and does not change the ultraviolet structure of the observables.

In order to completely fix the renormalisation scheme for quark masses,
we still need to provide a definition of the renormalised coupling.
This allows to relate the scale $\mu=1/L$ to the bare coupling, and hence
to the lattice spacing, in an unambiguous way, so that the continuum limit
of $\SigmaP$ is precisely defined.
Following~\cite{Brida:2016flw,DallaBrida:2016kgh}, we will introduce two different definitions, to be used
in qualitatively different regimes. For renormalisation scales larger
than some value $\muswi/2$, we will employ the non-perturbative
SF coupling first introduced in~\cite{Luscher:1992an}. Below that scale, we will use
the gradient flow (GF) coupling defined in~\cite{Fritzsch:2013je}. As discussed in~\cite{Ramos:2015dla},
this allows to optimally exploit the variance properties of the couplings,
so that a very precise computation of the $\beta$ function, and ultimately
of the $\lQCD$ parameter, is achieved.\footnote{Also here, both couplings are computed
from correlation functions projected to the $Q=0$ sector of the theory.}
In our context, the main consequence of this setup
is that our quark masses are implicitly defined in two different
schemes above and below $\muswi/2$;
we will refer to them as $\HEs$ and $\LEs$, respectively.
Note that the schemes differ only by the choice of renormalized coupling $\gbar^2$;
the definition of $\ZP$ is always given by Eq.~\eqref{eq:ZP}.

The value of $\mu_0$ is implicitly fixed by
\begin{gather}
\label{eq:switching_scale}
\gbar_{\rm\scriptscriptstyle SF}^2(\mu_0)=2.0120\,,
\end{gather}
where one has~\cite{Brida:2016flw}
\begin{gather}
\label{eq:switching_scale_matchSF}
\gbar^2_{\rm\scriptscriptstyle SF}(\muswi/2) = \sigma(2.0120) = 2.452(11)\,.
\end{gather}
The running of the SF coupling is thus known accurately down to
$\muswi/2$, and the matching of the two schemes is completely specified by measuring
the value of the GF coupling at $\muswi/2$, for which one has~\cite{DallaBrida:2016kgh}
\begin{gather}
\label{eq:switching_scale_match}
\gbar_{\rm\scriptscriptstyle GF}^2(\mu_0/2)=2.6723(64)\,.
\end{gather}
When expressed in physical units through the ratio $\muswi/\lQCD$, one
finds $\muswi\approx 4\,\GeV$~\cite{Bruno:2017gxd} --- i.e., the scheme switching 
takes place at a scale around $2~\GeV$.
It is important to stress that the scheme definition affects different quantities in distinct ways.
Obviously, the $\beta$ function, being a function of the coupling, will be different in the two schemes.
The same is true of the mass anomalous dimension $\tau(g)$.
The renormalised masses $\mbar_i(\mu)$, on the other hand,
are smooth functions across $\muswi/2$ by construction, since
--- unlike the RG functions $\beta$ and $\tau$, which are functions
of $g$ --- they are functions of the scale $\mu$, and are fixed by
the same definition of $\ZP$ at all scales.
This observation also provides the matching relation for the anomalous dimensions:
for any fixed $\mu$ we have
\begin{gather}
\tau_{\scriptscriptstyle\HEs}(\gbar_{\rm\scriptscriptstyle SF}^2(\mu)) =
\tau_{\scriptscriptstyle\LEs}(\gbar_{\rm\scriptscriptstyle GF}^2(\mu))\,.
\end{gather}

Another important motivation for this choice of strategy is the 
control over the perturbative expansion of the $\beta$ function and mass anomalous
dimension, which becomes relevant at very high energies. In the SF scheme
the first non-universal perturbative coefficient of the $\beta$ function, $b_2$,
is known~\cite{Bode:1999sm},
\begin{gather}
  \label{eq:b2sf}
b_2=\frac{1}{(4\pi)^3}\left(0.483-0.275\NF+0.0361\NF^2-0.00175\NF^3\right)\,.
\end{gather}
Moreover, the next-to-leading order (NLO) mass anomalous
dimension in the $\HEs$ scheme was computed in~\cite{Sint:1998iq},
\begin{gather}
d_1 = \frac{1}{(4\pi)^2}\left(0.2168+0.084\NF\right)\,.
\end{gather}
A similar computation in the $\LEs$ scheme is currently not available, due to
the absence of a full two-loop computation of the finite-volume GF coupling
in QCD.

Let us end this section summarising the results for the $\beta$ function in our
choice of schemes. As discussed above, these results will be essential
to the determination of the anomalous dimension $\tau(\gbar)$ in the following
sections. On the high-energy side we have~\cite{Brida:2016flw}
\begin{align}\label{eq:betasf}
   \beta_{\rm\scriptscriptstyle SF}(\gbar) &= -\gbar^3\sum_{n=0}^3 b_n\gbar^{2n}\,, & \gbar^2 &\in [0,2.45] \,,
\end{align}
with $b_{0,1}$ given by Eq.~\eqref{eq:univ}, $b_2$ given by
Eq.~\eqref{eq:b2sf}, and $b_3=b_3^{\rm eff}$ a fit parameter with value
\begin{equation}
  (4\pi)^4 b_3^{\rm eff} = 4(3)\,.
\end{equation}
Note that the three leading coefficients are given by the perturbative
predictions, which implies that safe contact with the asymptotic
perturbative behavior has been made (this is the reason why
Eq.~\eqref{eq:betasf} is accepted as a reliable approximation all the way up to $\gbar=0$). 
On the
other hand, on the low energy side, we have~\cite{DallaBrida:2016kgh} 
\begin{subequations}
  \label{eq:betagf}
\begin{align}
  \beta_{\rm\scriptscriptstyle GF}(\gbar) &= -\frac{\gbar^3}{\sum_{n=0}^2 p_n\gbar^{2n}} \,, &
   \qquad \gbar^2 &\in [2.1,11.3]\,,
\end{align}
with fit parameters
\begin{equation}
  \label{eq:gfparm}
  p_0 = 16.07\,,\quad p_1 = 0.21\,,\quad p_2=-0.013\,,
\end{equation}
and covariance matrix 
\begin{equation}
  \label{eq:gfcov}
  {\rm cov}(p_i,p_j) = \left(
    \begin{array}{rrr}
      5.12310\times 10^{-1}& -1.77401\times 10^{-1}& 1.32026\times 10^{-2}\\
      -1.77401\times 10^{-1}& 6.60392\times 10^{-2}& -5.10305\times 10^{-3}\\
      1.32026\times 10^{-2}& -5.10305\times 10^{-3}& 4.06114\times 10^{-4}\\
    \end{array}
\right)\,.
\end{equation}
\end{subequations}

The reader should note that these values are \emph{not} exactly the
same as those quoted as final results in~\cite{DallaBrida:2016kgh}. There are two
reasons for this. First we have added some statistics in some
ensembles, where the uncertainty in $\SigmaP$ was too large. Second, a
consistent treatment of the correlations and autocorrelations between
$\ZP$ and $\gbar_{\rm\scriptscriptstyle GF}^2$ requires knowledge of the joint
autocorrelation function in a consistent way. This
requirement results in a different covariance matrix between the fit
parameters. In any case it is very
important to point out that both results, Eqs.~(\ref{eq:gfparm},
\ref{eq:gfcov}), and those quoted in~\cite{DallaBrida:2016kgh} are perfectly
compatible. The reader can easily check that the
differences in the $\beta$ function are completely
negligible within the quoted uncertainties.

  All data is analysed using the $\Gamma$-method~\cite{Wolff:2003sm} to account
  for autocorrelations (some integrated autocorrelation times are given in the
  tables of appendix~\ref{app:sdetail}). For a full propagation of uncertainties
  into derived quantities, we subsequently apply standard resampling techniques (boostrap).

\subsection{Determination of RGI quark masses}

In order to determine RGI quark masses, we will factor Eq.~(\ref{eq:mrgidef}) as
\begin{gather}
  \label{eq:factors}
\mrgi_i = \frac{\mrgi_i}{\mbar_i(\mupt)}\,
\frac{\mbar_i(\mupt)}{\mbar_i(\muswi/2)}\,
\frac{\mbar_i(\muswi/2)}{\mbar_i(\muhad)}\,
\mbar_i(\muhad)\,.
\end{gather}
The three ratios appearing in this expression are flavour-independent
running factors:\footnote{The relevant quark masses $\mbar_i$ are always
renormalised as in \req{eq:ZP}. This is usually called the SF renormalisation scheme
but, as previously explained, in the present work we employ SF and GF renormalisation
conditions for the gauge coupling. We use $\HEs$ and $\LEs$ to also label our quark mass anomalous 
dimensions.}
\begin{itemize}
\item $\mbar_i(\muswi/2)/\mbar_i(\muhad)$ is the running between
some low-energy scale $\muhad$ and the scheme-switching scale $\muswi/2$.
It will be computed non-perturbatively in the $\LEs$ scheme.
\item $\mbar_i(\mupt)/\mbar_i(\muswi/2)$ is the running between
the scheme-switching scale $\muswi/2$ and some high-energy scale $\mupt$.
It will be computed non-perturbatively in the $\HEs$ scheme.
\item $\mrgi_i/\mbar_i(\mupt)$ can be computed perturbatively
using NLO perturbation theory in the $\HEs$ scheme.
This perturbative matching would be safe, entailing a small systematic uncertainty
due to perturbative truncation, provided $\mupt$ is large enough
--- say, $\mupt$ of order $M_W$. As discussed in Sec.~\ref{sec:sf},
we will actually use our non-perturbative results for the
mass anomalous dimension at high energies to constrain the truncation
systematics, and obtain a very precise matching to perturbation theory.
\end{itemize}
Finally, the renormalised quark mass $\mbar_i(\muhad)$ at the low-energy scale
is to be computed independently from the running factors, by determining bare PCAC
quark masses $m_i$ from large-volume lattice simulations at a number of values of the lattice
spacing $a$ --- or, equivalently, of the bare lattice coupling $g_0^2$ ---
and combining them with suitable $\LEs$ scheme renormalisation factors $\Zm$ as
\begin{gather}
\label{eq:Zhad}
\mbar_i(\muhad) = \lim_{a \to 0} \Zm(g_0^2,a\muhad)\,m_i(g_0^2)\,.
\end{gather}
Therefore, the complete renormalisation programme for light quark
masses requires 
the computation of each of the three running factors to high
precision, as well 
as the determination of $\Zm$ for the regularisation eventually
employed in the computation 
of $m_i(g_0^2)$, within the appropriate range of values of $g_0^2$.

\section{Running in the high-energy region}
\label{sec:sf}

\subsection{Determination of $\ZP$ and $\SigmaP$}

Our simulations in the high-energy range above $\muswi$ have been
performed at eight different values of the renormalised Schr\"odinger
Functional coupling 
\begin{equation}
  \uSF \in \{1.1100, 1.1844, 1.2656, 1.3627, 1.4808, 1.6173, 1.7943,
2.0120\}\,,
\end{equation}
for which we have determined the step scaling function $\SigmaP$ of
Eq.~(\ref{eq:ssfcl}) 
at three different values of the lattice spacing $L/a=6,8,12$. 
At the strongest
coupling $\uSF=2.012$ we have also simulated an extra finer
lattice with $L/a=16$, in order to have a stronger crosscheck
of our control over continuum limit extrapolations in the less
favourable case.
The value of the hopping
parameter $\kappa$ is tuned to its critical value $\hopc$, so that the
bare $\Oa$-improved PCAC mass
\begin{gather}
\label{eq:mpcac} m(g_0^2,\kappa) =
\left.\frac{\half(\partial_0^*+\partial_0)\fA(x_0)+\icA
a\partial_0^*\partial_0\fP(x_0)}{2\fP(x_0)}\right|_{x_0=T/2}\,,
\end{gather} vanishes\footnote{Details can be found
in~\cite{Nf3tuning}; a discussion of the systematic impact of this procedure on
our data is provided in Appendix~\ref{app:sys}.}
at the corresponding value of
$\beta=6/g_0^2$.
Simulations have been carried out using the plaquette gauge action~\cite{Wilson:1974sk},
and an $\Oa$-improved fermion action~\cite{Sheikholeslami:1985ij} with the
non-perturbative value of the improvement coefficient $\icsw$~\cite{Yamada:2004ja},
and the one-loop~\cite{Luscher:1996vw} and two-loop~\cite{Bode:1998hd} values, respectively,
of the boundary improvement counterterms $\icttil$ and $\ict$.
All the simulations in this paper were performed using
a variant of the \texttt{openQCD} code~\cite{Luscher:2012av,openqcd}.

The data for $\SigmaP$ can be corrected by subtracting the cutoff
effects to all orders in $a$ and leading order in $g_0^2$, using the
one-loop computation of~\cite{Sint:1998iq}, viz. 
\begin{align}\label{eq:1lc}
   \SigmaPI(u,a/L) &= \frac{\SigmaP(u,a/L)}{1 + u \deltaP(a/L)} \,, &
      \deltaP(a/L) &= -d_0\ln(2)c(L/a) \,,
\end{align}
where $c(a/L)$ is given in Table~\ref{tab:1limp}. This correction
is bigger than our statistical uncertainties for $L/a=6$,
but smaller than the ones for $L/a>6$. As will be discussed below,
the scaling properties of $\SigmaPI$ are somewhat
better than those of the unsubtracted $\SigmaP$ --- and, more importantly,
the study of the impact of the perturbative subtraction will allow us
to assign a solid systematic uncertainty related to the continuum
limit extrapolation.

The results of our simulations are summarised in
Table~\ref{tab:ZP_SF}. Alongside the results for $\ZP$ at each
simulation point, we quote the corresponding values of
$\SigmaPI$. The first uncertainty is statistical,
while the second one is an estimate of the systematic uncertainty
due to $\mbox{O}(a^{n>2})$ cutoff effects, given by the difference
of the one-loop corrected and uncorrected values of the SSF,
$\SigmaPI-\SigmaP$.

\begin{table}[t!]
    \centering
    \begin{tabular}{RCRC}\toprule
     L/a & c(a/L)    & L/a & c(a/L)    \\\cmidrule(lr){1-2}\cmidrule(lr){3-4}
      6  & +0.020787 & 16  & -0.005210 \\
      8  & -0.002626 & 18  & -0.004605 \\
      10 & -0.006178 & 20  & -0.004073 \\
      12 & -0.006368 & 22  & -0.003615 \\
      14 & -0.005848 & 24  & -0.003224 \\
    \bottomrule
    \end{tabular}
    \caption{One-loop cutoff effects in $\SigmaP$ in the $\HEs$ scheme.
            }
    \label{tab:1limp}
\end{table}

\subsection{Determination of the anomalous dimension}

Once the lattice step scaling function $\SigmaP(u,a/L)$ is known, 
we are in a position to determine the RG running of the light quark masses
between the hadronic and electro-weak energy scales. This we do using
four methods; though equivalent in theory, they are distinct numerical
procedures. Thus they give us insight into the magnitude of the systematic
errors of our results. Two of these procedures consist in extrapolations of
$\SigmaP(u,a/L)$  to the continuum limit. Knowledge of the continuum SSF
$\sigmaP(u)$ is adequate for controlling the RG evolution between
energy scales. The other two procedures essentially extract the quark mass
anomalous dimension $\tau$ from $\sigmaP$, using Eq.~\eqref{eq:sigmaP}.
In this way we have multiple crosschecks on the final result.

\begin{figure}[t!]
  \centering
  \includegraphics[width=0.7\textwidth]{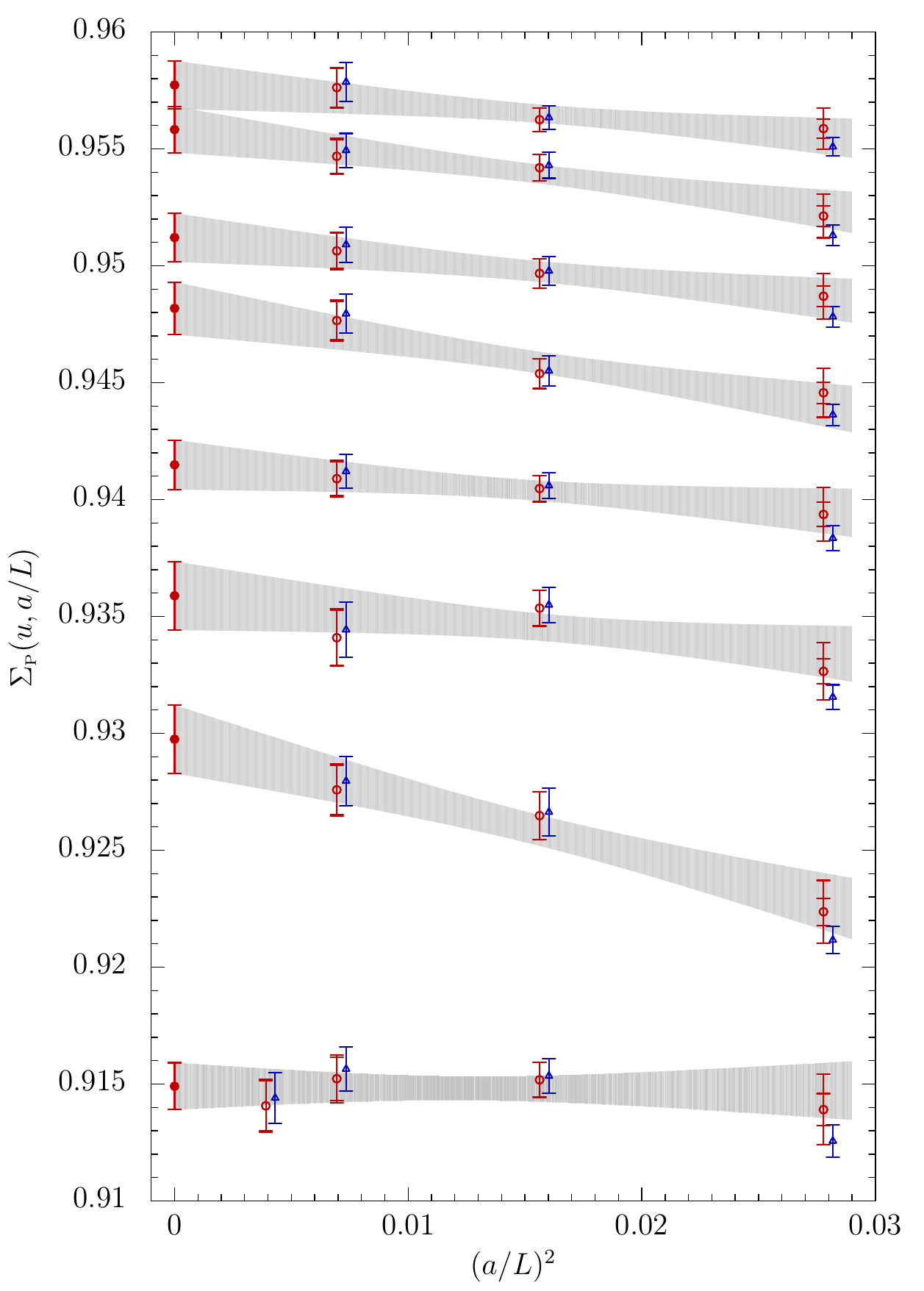}
  \caption{Continuum extrapolations of $\SigmaP$ at fixed value of $u$ in the
  high-energy region. Open blue triangles correspond to $\SigmaP$, while open
  red circles correspond to $\SigmaPI$; the latter include the systematic uncertainty
  discussed in the text, visible as a double error bar only in the $L/a=6$ data.
  The extrapolations shown (filled red circles and grey bands) are those for $\SigmaPI$.}
  \label{fig:contSF}
\end{figure}

The first procedure is labelled as $\sigmaP$:\emph{u-by-u}. It
starts with the continuum extrapolation of
$\SigmaPI(u,a/L)$ at fixed $u$, using the ansatz
\begin{equation}
\label{eq:SigmaP-extrap}
  \SigmaPI(u, a/L) = \sigmaP(u) + \rho_{\rm\scriptscriptstyle P}(u)\left(\frac{a}{L}\right)^2\, .
\end{equation}
With $u$ held constant, $\sigmaP$ and $\rho_{\rm\scriptscriptstyle P}$ are fit
parameters. A detailed study shows 
that when the data for $\SigmaPI(u, a/L)$ are extrapolated
to the continuum linearly in $(a/L)^2$, the effect of the subtraction
of cutoff effects at one-loop becomes noticeable, cf. Fig.~\ref{fig:contSF}.
The fits to the unsubtracted values of $\SigmaP$ have a total $\chi^2/\text{dof} = 12.9/9$, while the
fits to the subtracted data have $\chi^2/\text{dof} = 8.6/9$ (with ``total" above meaning $\chi^2/\text{dof}$,
summed over all fits).
Based on this observation, we add the systematic uncertainty
quoted in Table~\ref{tab:ZP_SF} in quadrature to the statistical
uncertainty of $\SigmaPI$, and use the result as input
for our fits. This procedure increases the size of the uncertainties
of our continuum-extrapolated values by about a 20-30\%. Table~\ref{tab:ZP_SF}
quotes $\sigmaP(u)$ results at the eight values of the coupling,
as well as the respective slopes $\rho_{\rm\scriptscriptstyle P}$,
from this conservative analysis. 

The next step in this procedure consists in fitting the eight extrapolated
$\sigmaP(u)$ results to a polynomial 
of the form
\begin{equation}
\label{eq:sigmaP-series}
  \sigmaP(u) = 1 + s_1 u + \sum_{n=2}^{n_s} c_n u^n\,,
\end{equation}
so as to have a continuous expression for $\sigmaP(u)$.
The leading non-trivial coefficient is always set
to the perturbative universal prediction $s_1 = -d_0\ln(2)$.
The $O(u^2)$ parameter can either be left as a free fit parameter
or held fixed to the perturbative
value $c_2=s_2=-d_1\ln(2)+(\half d_0^2-b_0d_0)\ln(2)^2$.
Higher-order coefficients $c_{n > 2}$ are free fit parameters.
We label as \texttt{FITA} the one with a free $c_2$ and
as \texttt{FITB} the one with  $c_2=s_2$.
The series expansion of \req{eq:sigmaP-series} is truncated
either at $n_s = 4$ or $n_s = 5$.

Finally, the resulting continuous function for $\sigmaP(u)$
is readily calculated for the coupling values provided by the recursion 
\begin{gather}
  \gbar^2_{\rm\scriptscriptstyle SF}(\muswi) = 2.012\,, \qquad u_k = \gbar^2_{\rm\scriptscriptstyle SF}(2^k\muswi) \, ,
  \label{eq:ukrec}
\end{gather}
and the RG evolution can be determined by the renormalised-mass ratios
at different scales
\begin{gather}
\begin{split}
\label{eq:Rk-sigma}
  R^{(k)} & 
   \equiv \frac{\mbar(2^k\muswi)}{\mbar(\muswi/2)} = \prod_{n=0}^k \sigmaP(u_n)\,,
\end{split}
\end{gather}
cf. \req{eq:sigmaalt}.
Note that this procedure is the one previously employed for the determination
of the running in the $\NF=0$~\cite{Capitani:1998mq} and $\NF=2$~\cite{DellaMorte:2005kg} cases.

Our second procedure, labelled as $\sigmaP$:\emph{global},
is a global analysis of our data, in which the $\SigmaPI(u,a/L)$-extrapolation
is performed with respect to {\it both} variables $u$ and $a/L$. 
We extrapolate our dataset using \req{eq:SigmaP-extrap},
with $\sigmaP(u)$ expanded according to \req{eq:sigmaP-series},
and $\rho_{\rm\scriptscriptstyle P}(u)$ 
expanded according to
\begin{equation}
  \rho_{\rm\scriptscriptstyle P}(u) = \sum_{n=2}^{n_\rho} \rho_n u^n\,.
\label{eq:rho-series}
\end{equation}
Since our data have been corrected for cutoff effects up to one-loop,
we consistently drop terms $u^0$ and $u^1$ from the above 
polynomial. This series expansion is truncated
at either $n_\rho =2$ or $n_\rho = 3$. The series expansion 
of \req{eq:sigmaP-series}, is truncated either at $n_s = 4$ or $n_s = 5$.
Again, the choice of $c_2$  is labelled as
\texttt{FITA} (if it is left as a free fit parameter) or \texttt{FITB} (if $c_2=s_2$). 
Once $\sigmaP(u)$ has been obtained from the
global extrapolation, the RG running is determined from \req{eq:Rk-sigma}, just like
in procedure $\sigmaP$:\emph{u-by-u}.

The third procedure, labelled as $\tau$:\emph{u-by-u}, starts off just like
$\sigmaP$:\emph{u-by-u}: at constant $u$, we fit the datapoints $\SigmaPI(u, a/L)$
with \req{eq:SigmaP-extrap}, obtaining $\sigmaP(u)$. 
Then the continuum values $\sigmaP(u)$ are fit with Eq.~\eqref{eq:sigmaP},
where in the integrand we use the results of Sect.~\ref{sec:str}
for the $\beta$ function (cf. Eqs.~(\ref{eq:betasf}-\ref{eq:betagf})) and the polynomial ansatz
\begin{equation}
  \tau(x) = -x^2\sum_{n=0}^{n_s} t_n x^{2n}
  \label{eq:tau-series}
\end{equation}
for the quark mass anomalous dimension. We fix the two leading coefficients to the
perturbative asymptotic predictions $t_0=d_0\approx 0.05066$ and $t_1=d_1\approx 0.002969$ 
(this is labelled as \texttt{FITB}).
The coefficients $t_{n>1}$ are free fit parameters and the series is truncated at $n_s = 2, \ldots, 5$.

Having thus obtained an estimate for the anomalous dimension $\tau(u)$, we  
arrive at another determination of the renormalised-mass ratios, using the expression
\begin{gather}
\label{eq:recursion2}
\begin{split}
  R^{(k)} & 
  \equiv \frac{\mbar(2^k\muswi)}{\mbar(\muswi/2)}
  = \exp\left\{-\int_{\sqrt{u_k}}^{\sqrt{u_{-1}}} \dif x\, \frac{\tau(x)}{\beta(x)} \right\} \,,
\end{split}
\end{gather}
with the couplings determined through \req{eq:ukrec}.

Our fourth procedure, labelled as $\tau$:\emph{global}, consists in
performing the continuum extrapolation of $\SigmaP(u,a/L)$ and the
determination of the anomalous dimension $\tau(u)$ simultaneously. Once
more, $\SigmaP(u,a/L)$ is treated as a function of two variables.
This approach is based on the relation 
\begin{equation}
  \ln\left(\SigmaPI(u,a/L) - \rho_{\rm\scriptscriptstyle P}(u)\left(\frac{a}{L}\right)^2\right) = 
  - \int_{\sqrt{u}}^{\sqrt{\sigma(u)}} \dif x\, \frac{\tau(x)}{\beta(x)}\,.
\end{equation}
with $\rho_{\rm\scriptscriptstyle P}(u)$ and $\tau(u)$ parameterised by the
polynomial ans\"atze (\ref{eq:rho-series}) and (\ref{eq:tau-series}) respectively, and
the $\beta$ function being again provided by Eqs.~(\ref{eq:betasf}-\ref{eq:betagf}).
In practice the $\rho_{\rm\scriptscriptstyle P}(u)$-series is truncated at $n_\rho = 2,3$.
Again we account for the one-loop correction of our data for cutoff effects by
consistently dropping terms $u^0$ and $u^1$ from the 
$\rho_{\rm\scriptscriptstyle P}(u)$-polynomial.\footnote{We have also fit the unsubtracted 
$\SigmaP(u,a/L)$ with a similar global fit approach, obtaining
compatible results. We note in passing that in this case consistency requires that only the  $u^0$ term
is dropped in \req{eq:rho-series}.}
For the $\tau$-series, we always fix the leading universal coefficient to the
perturbative asymptotic prediction $t_0=d_0\approx 0.05066$, while
we either leave the rest of the coefficients to be determined
by the fit (labelled \texttt{FITA}), or fix the ${\rm O}(x^4)$ coefficient to the 
perturbative prediction $t_1=d_1\approx 0.002969$ (labelled \texttt{FITB}).

Like in the previous  $\tau$:\emph{u-by-u} procedure,
having obtained an estimate for the anomalous dimension $\tau(u)$, we again 
arrive at an expression for the renormalised-mass ratios using \req{eq:recursion2}.

The main advantage of the two \emph{u-by-u} analyses
is that one has full control over the continuum
extrapolations, since they are performed independently of the
determination of the anomalous dimension. The disadvantage is that
having to fit, for most $u$-values, three datapoints  with the two-parameter
function (\ref{eq:SigmaP-extrap}), we are forced to include our $L/a=6$ results,
which are affected by the largest discretisation errors. As far as the two \emph{global}
analyses are concerned, they have two advantages. First, one can
choose not to include the coarser lattice data points, i.e., one can
fit only to the data with $L/a>6$. This provides an extra handle on the
control of cutoff effects. Second, slight mistunings in the value
of the coupling $u$ can be incorporated by the fit.

In order to have a meaningful quantitative comparison of the
four procedures, we display in Table~\ref{tab:mrat} results for
$R^{(k)}$, obtained from all four methods and for
a variety of fit ans\"atze. In general,
the agreement is good, though it is clear that the fit quality improves
when the data with $L/a=6$ are discarded.
Fits for $\tau$ that do
not use the known value for $t_1$ tend to have larger errors, as
expected, since the asymptotic perturbative behavior is not constrained
by the known analytic results.
We therefore focus on \texttt{FITB}.
Concerning fits from procedures $\sigmaP$:\emph{u-by-u}
and $\sigmaP$:\emph{global}, the parameterisation with $n_s=5$
tends to provide a better description of our data.
Anyway, the key point is that the 
analysis coming from the recursion of the step scaling functions is in good
agreement with that coming from a direct determination of the anomalous dimension.

\begin{figure}[t!]
  \centering
  \includegraphics[width=0.9\textwidth]{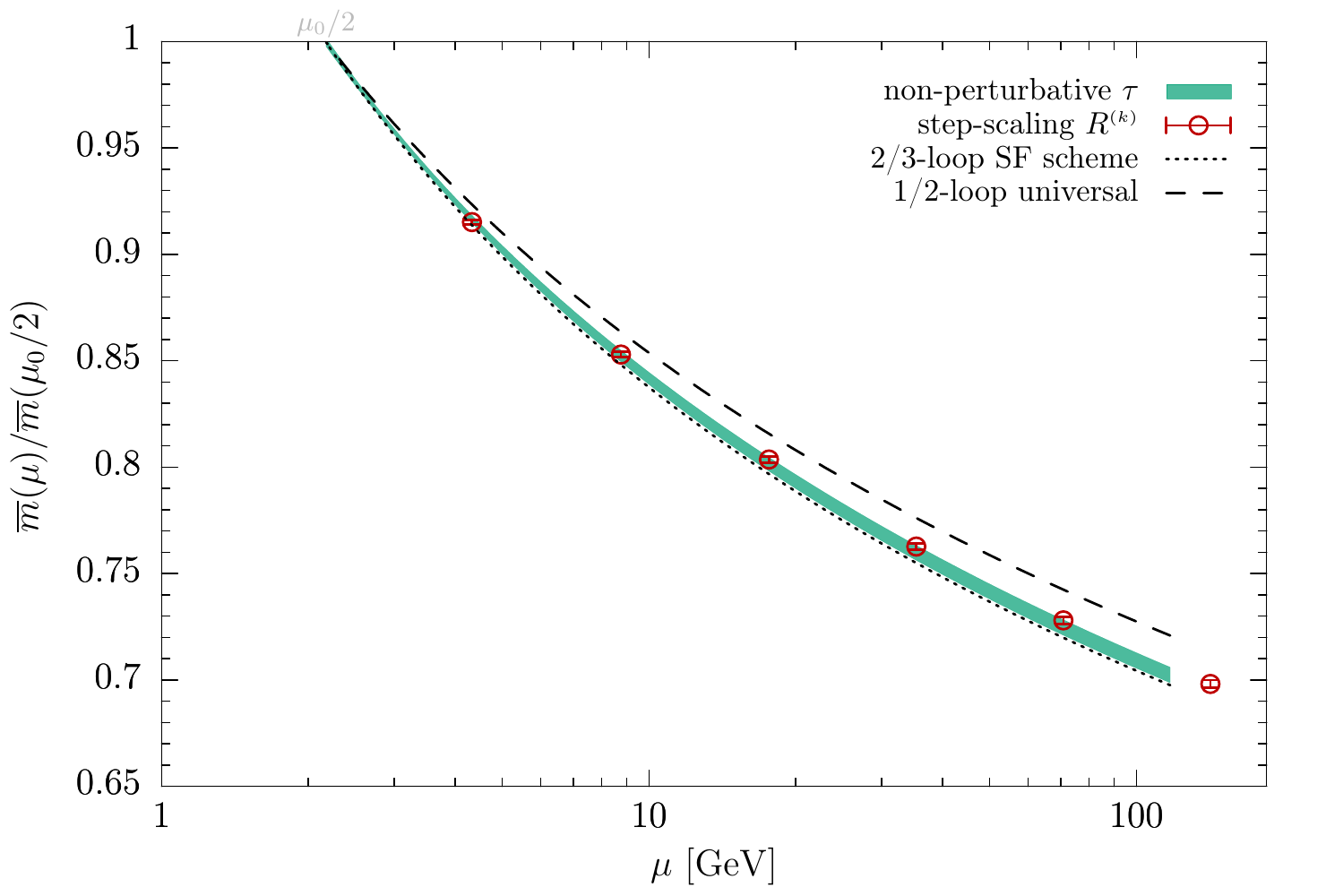}
  \caption{RG evolution factor towards high energies from the lowest energy
           scale $\muswi/2$ reached with the $\HEs$ scheme, given by the ratio
           $\mbar(\mu)/\mbar(\muswi/2)$.  The green band is the result from our
           preferred determination of the anomalous dimension, Eq.~\eqref{eq:FITB*}, 
           in the region covered by our data, while the red
           points are the values of $R^{(k)}$ obtained from the step scaling in
           factors of $2$ based on $\sigmaP$. For the latter we have used the
           \emph{u-by-u} \texttt{FITB} results with $n_s=5$ from
           Table~\ref{tab:mrat}. The scale setting to obtain $\mu$ in physical
           units uses $\muswi\approx 4\,\GeV$, obtained from
           $\lQCD$ in~\cite{Bruno:2017gxd}.  Perturbative predictions at
           different orders are also shown for comparison.
	          }
  \label{fig:running_sf}
\end{figure}

Based on this discussion, we quote as our preferred result the determination
of $\tau$ from a global \texttt{FITB} without the $L/a=6$ data, and $n_s=n_\rho=2$.
We refer to this determination as \texttt{FITB*} in the following. 
The  result for the anomalous dimension at high energies is thus given by
\begin{gather}
\nonumber
\tau(g) = -g^2 \sum_{n=0}^2 t_n g^{2n}\,;\\ \label{eq:FITB*}
  t_0 = \frac{8}{(4\pi)^2}\,, \qquad
  t_1 = 0.002969\,, \qquad
  t_2 = -0.00009(26)\,,
\end{gather}
and is illustrated in Fig.~\ref{fig:tau}.
We stress that this result comes from a conservative approach, since
we drop the $L/a=6$ data, and the statistical error of the points
is increased by the value of the one-loop prediction for cutoff effects.
To illustrate the good agreement of the various determinations
of the running mass, Fig.~\ref{fig:running_sf} illustrates the comparison between
our preferred fit for the anomalous dimension and the values obtained
from the recursion based on $\sigmaP$, demonstrating the excellent
level of agreement between otherwise fairly different analyses,
as well as the comparison with perturbative predictions.

\subsection{Connection to RGI masses}\label{sec:RGIhe}

In order to make the connection with RGI masses, as spelled out in our strategy
in Sec.~\ref{sec:str}, we could apply NLO perturbation theory directly
at an energy scale $\mu_{\rm\scriptscriptstyle pt}$ at the higher end of
our data-covered range --- e.g., the one defined by
$\gbar_{\rm\scriptscriptstyle SF}^2(\mu_{\rm\scriptscriptstyle pt})=1.11$ ---
to compute $M/\mbar(\mu_{\rm\scriptscriptstyle pt})$, and multiply it with the
non-perturbative result for $\mbar(\mu_{\rm\scriptscriptstyle pt})/\mbar(\muswi/2)$.
We however observe that the perturbative description of the running above $\mu_{\rm\scriptscriptstyle pt}$ can
be constrained by employing our non-perturbatively determined form
of $\tau$, which at very small values of the coupling agrees with the
asymptotic perturbative expression by construction.
It is thus possible to directly compute the quantity
\begin{eqnarray}
  \label{eq:Movm}
  \frac{M}{\mbar(\muswi/2)} &=& 
  [2b_0\gbar_{\rm\scriptscriptstyle SF}^2(\muswi/2)]^{-d_0/2b_0}
  \exp\left\{ 
    -\int_0^{\gbar_{\rm\scriptscriptstyle SF}(\muswi/2)}\kern-15pt \dif x \left[ \frac{\tau(x)}{\beta(x)} -
      \frac{d_0}{b_0x}\right] 
  \right\}\,,
\end{eqnarray}
using as input the $\tau$ function from different fits, in order to
assess the systematic uncertainty of the procedure.

\begin{table}
  \centering
  \begin{tabular}{>{\ttfamily}lcccccc}\toprule
    Type & $n_s$ & $n_\rho$ & $[L/a]_{\rm min}$& $M/m(\mu_0/2)$& $\chi^2/{\rm dof}$ \\\midrule
    FITB & 2 & 2 & 6& 1.7577(77)& 18/23 \\ 
    FITB*& 2 & 2 & 8& 1.7505(89)& 12/15 \\ 
    FITB & 3 & 2 & 6& 1.7580(80)& 18/22 \\ 
    FITB & 3 & 2 & 8& 1.7500(97)& 12/14 \\ 
    \bottomrule
  \end{tabular}
  \caption{Various fits of the anomalous dimension and results for
           $M/\mbar(\mu_0/2)$, see section~\ref{sec:RGIhe}.
          }
  \label{tab:Movm}
\end{table}

The result of this exercise for a selection of global fits for $\tau$,
both with and without the $L/a=6$ data, is provided in 
Table~\ref{tab:Movm}. The agreement among different
parameterisations of the anomalous dimension is very good.
The value coming from our preferred fit is
\begin{equation}
  \label{eq:MovmSF}
  \frac{M}{\mbar(\muswi/2)} = 1.7505(89)\,.
\end{equation}
This is our final result coming from the high-energy region,
bearing a $0.5\%$ error.
As stressed above, our error estimates can be deemed conservative;
an extra-conservative error estimate could be obtained
by adding in quadrature the maximum spread of central values
in Table~\ref{tab:Movm}, which would yield a $0.7\%$ final uncertainty.
We however consider the latter an overestimate, and stick
to Eq.~(\ref{eq:MovmSF}) as our preferred value.

\section{Running in the low-energy region}
\label{sec:gf}

As already explained, at energies $\mu < \mu_0/2$ it is
convenient to change to the GF scheme. The objective of the low-energy
running is to compute the ratio $\mbar(\mu_0/2)/\mbar(\muhad)$, that,
together with the ratio $M/\mbar(\mu_0/2)$ of eq.~\eqref{eq:Movm},
will provide the total running factor.

We have again determined the
step scaling function $\SigmaP$ of Eq.~(\ref{eq:ssfcl}) at three
different values of the lattice spacing, now using lattices of size
$T/a=L/a=8,12,16$, and double lattices of size $L/a=16,24,32$.  The
bare parameters are chosen so that $\uGF$ 
is approximately equal to one of the seven values
\begin{equation}
  \{2.12, 2.39, 2.73, 3.20, 3.86, 4.49, 5.29\}\,.
\end{equation}
Note that the lattice sizes are larger than in the high-energy region. 
This allows to better tackle cutoff effects, which are expected to be 
larger because of the stronger coupling, and the larger scaling
violations exhibited by $\uGF$ with respect to $\uSF$~\cite{Fritzsch:2013je,Brida:2016flw,DallaBrida:2016kgh}.
Simulations have been carried out using a tree-level Symanzik improved (L\"uscher-Weisz)
gauge action~\cite{Luscher:1985zq}, and an $\Oa$-improved fermion action~\cite{Sheikholeslami:1985ij} with the
non-perturbative value of the improvement coefficient $\icsw$~\cite{Bulava:2013cta} and
one-loop values of the coefficients $\ict,\icttil$ for boundary
improvement counterterms~\cite{Takeda:2003he,Vilaseca,Nf3tuning}.  The chiral point is set using the same
strategy as before, cf.~Section~\ref{sec:sf}.
In contrast to the high-energy region, here the coupling constant
$\gbar_{\rm\scriptscriptstyle GF}^2$ and $\ZP$
are measured on the same ensembles, and hereafter we take the resulting
correlations into account in our analysis. 
As discussed in Section~\ref{sec:str}, computations are carried out at
fixed topological charge $Q=0$.  The main motivation for this is the
onset of topological freezing~\cite{Schaefer:2010hu} within the range of bare
coupling values covered by our simulations; setting $Q=0$ allows to
circumvent the large computational cost required to allow the charge
to fluctuate in the ensembles involved. The projection is implemented
as proposed in~\cite{Fritzsch:2013yxa,DallaBrida:2016kgh}. In practice, this is only an issue for the
finest ensembles at the largest values of $\uGF$ --- for $\uGF
\lesssim 4$ no configurations with nonzero charge have been
observed in simulations where the projection is not implemented.\footnote{It
is noteworthy in this context that in~\cite{Bulava:2015bxa,Bulava:2016ktf}
the improvement coefficient $\icA$ and the
normalisation parameter $\ZA$ of the axial current have been measured both for
a freely varying $Q$ and in the $Q=0$ sector, for several values of the bare gauge
coupling in the range listed in \ret{tab:hadmatch}. Results from both definitions were
found to be compatible.}

The results of our simulations are summarised in
Table~\ref{tab:ZP_GF}. Note that, contrary to the high-energy region,
here the value of $\uGF$ is not exactly tuned to a constant value
for different $L/a$. In practice, the slight mistuning is not visible
when extrapolating our data to the continuum, but our data set simply
begs to be analysed using the global approach described in
Section~\ref{sec:sf}. This approach only requires to have pairs of
lattices of sizes $L/a$ and $2L/a$ simulated at the same values of the
bare parameters. 

Moreover, there is no guarantee that when computing  
$\mbar(\mu_0/2)/\mbar(\muhad)$ the scale factor $\muhad/\mu_0$ is an
integer power of two. This speaks in favour of performing the analysis 
using the anomalous dimension. As in the previous section, we will
use the available information on the $\beta$ function,
eq.~\eqref{eq:betagf}. We will parameterise the ratio of RG functions as
\begin{equation}
  f(x) = \frac{\tau(x)}{\beta(x)} = \frac{1}{x}\sum_{n=0}^{n_r}f_n x^{2n}\,,
\end{equation}
and determine the fit parameters $f_n$ via a fit to the usual relation
\begin{equation}
  \log\left[ \SigmaP(u,a/L) - \rho(u)(a/L)^2 \right] =
  -\int_{\sqrt{u}}^{\sqrt{\sigma(u)}} \dif x\, f(x)\,.
\end{equation}
Once more, $\rho(u)$ describes the cutoff effects in $\sigmaP(u)$. 
When the fit parameters $f_n$ are determined, we can reconstruct the
anomalous dimension thanks to the relation
\begin{equation}
\label{eq:taugf}
  f(\gbar) = \frac{\tau(\gbar)}{\beta(\gbar)} 
  \quad\Longrightarrow\quad
  \tau(\gbar) = -\gbar^2 \frac{\sum_{n=0}^{n_r} f_n\gbar ^{2n}}
  {\sum_{k=0}^{k_t} p_k\gbar ^{2k}}\,.
\end{equation}
Recall that the parameters $p_n$ define our $\beta$ function in
Eq.~\eqref{eq:betagf}.

\begin{table}
  \centering
  \begin{tabular}{rcccc}\toprule
    &\multicolumn{2}{c}{$n_r=2$} & \multicolumn{2}{c}{$n_r=3$} \\\cmidrule(lr){2-3}\cmidrule(lr){4-5}
    $\rho(u)$ & $\mbar(\mu_0/2)/\mbar(\mu_{\rm had})$& $\chi^2/{\rm dof}$ &
    $\mbar(\mu_0/2)/\mbar(\mu_{\rm had})$& $\chi^2/{\rm dof}$\\\midrule
    $\rho_1u+\rho_2u^2$ & 0.5245(41) & 11/16 & 0.5250(42) & 11/15 \\
    $\rho_0 + \rho_1u+\rho_2u^2$  & 0.5222(43) & 9.5/15 & 0.5226(43) & 6.5/14 \\
    $\rho_1u+\rho_2u^2+ \rho_3u^3$ & 0.5201(45) & 7.4/14 & 0.5208(45)
                                                             & 6.0/13
    \\
    \bottomrule
  \end{tabular}
  \caption{Results for the running factor $\mbar(\mu_0/2)/\mbar(\mu_{\rm had})$}
  \label{tab:he_result}
\end{table}

\begin{figure}[t!]
  \centering
  \includegraphics[width=0.9\textwidth]{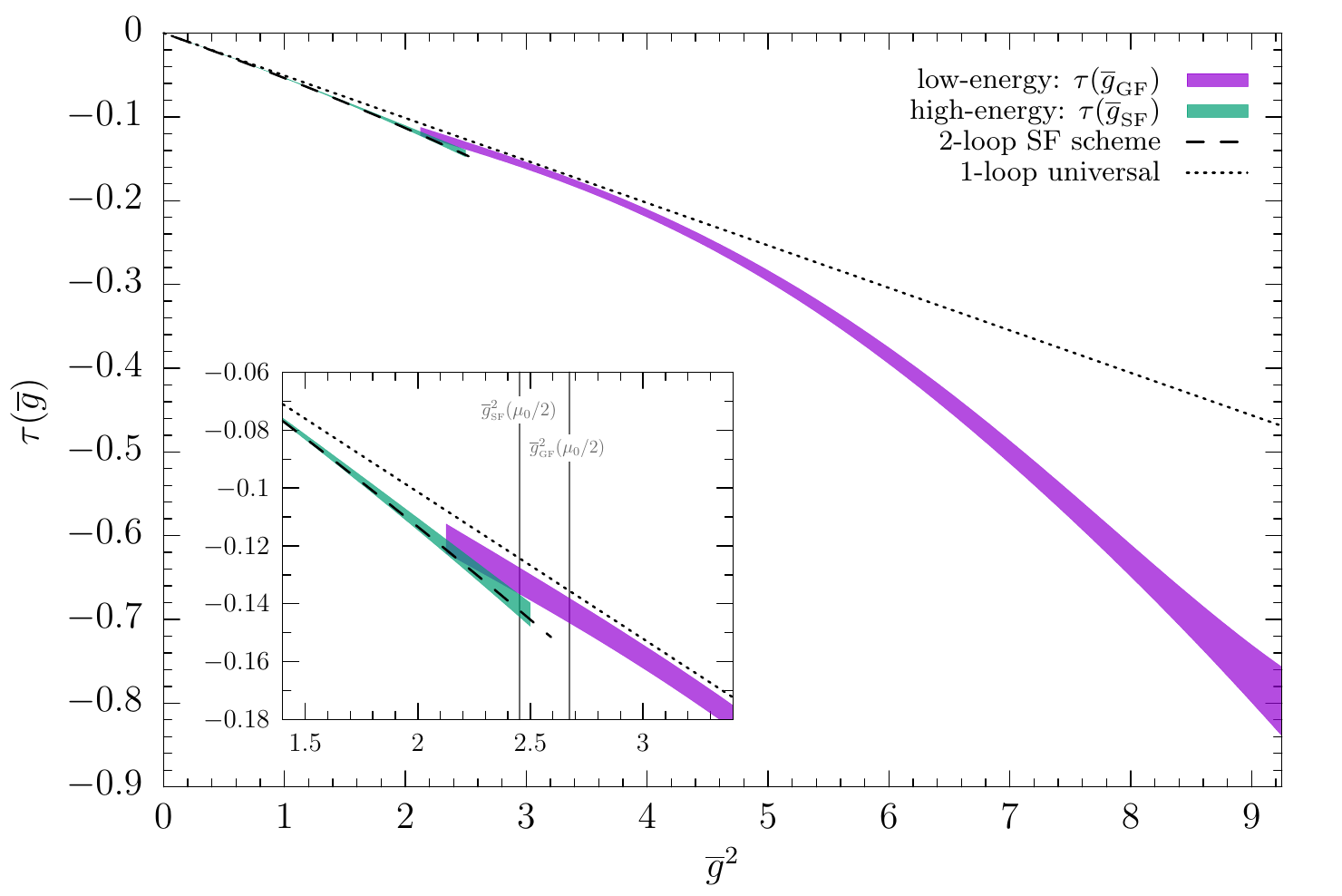}
  \caption{Non-perturbative mass anomalous dimension in the $\HEs$ (green) and $\LEs$ (purple)
  regions. Perturbative predictions at the highest available orders are also shown for comparison.}
  \label{fig:tau}
\end{figure}

\begin{figure}[h!]
  \centering
  \includegraphics[width=0.9\textwidth]{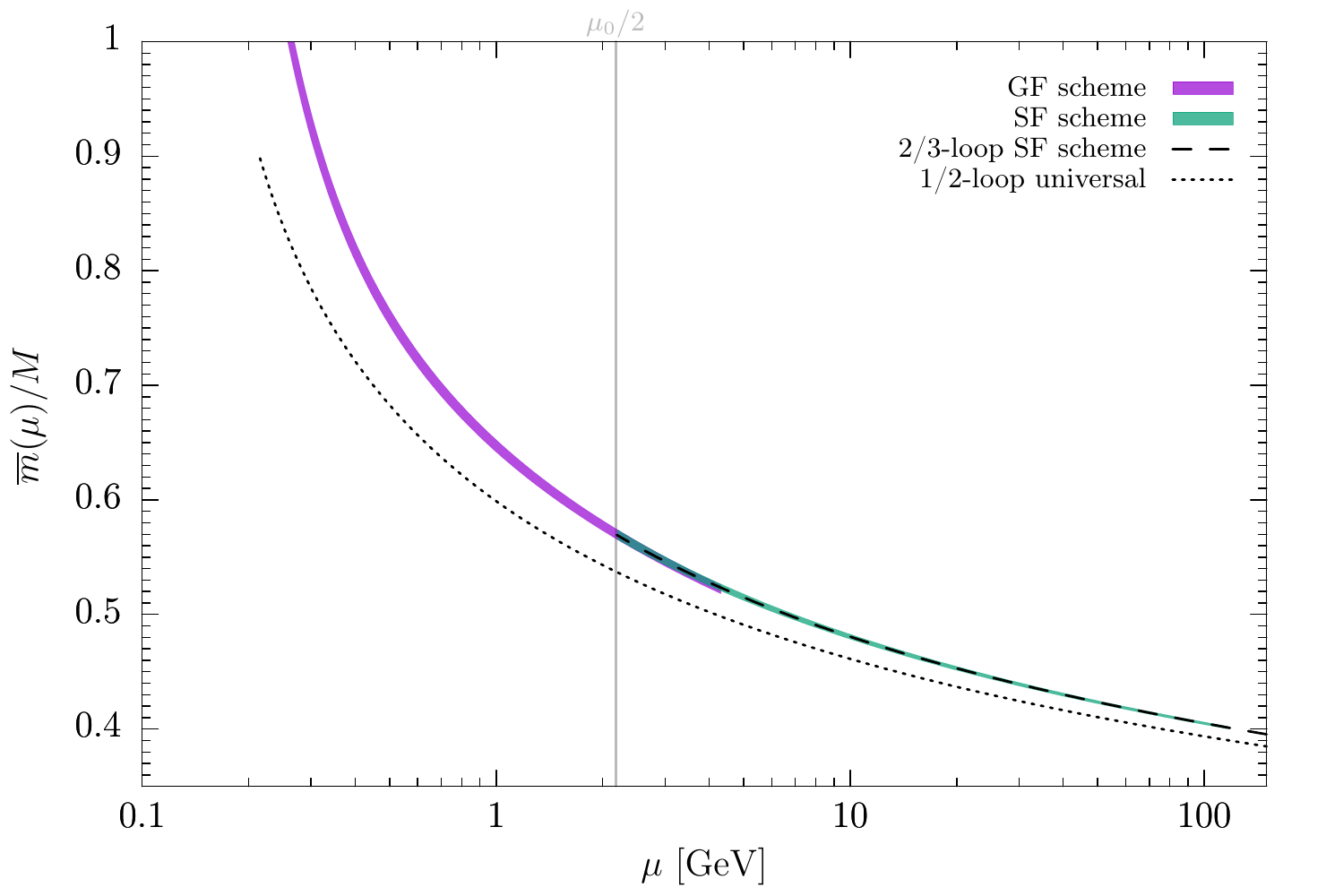}
  \caption{Non-perturbative running of light quark masses as a function of the
  energy scale, down to our hadronic matching scale. The value of $\mu$ in physical units has been obtained by
  using the result for $\lQCD$ in~\cite{Bruno:2017gxd}.
  Perturbative predictions at different orders are also shown for comparison.}
  \label{fig:runningmass}
\end{figure}

We have tried different ans\"atze, and as long as $n_r>1$ one gets a
good description of the data. The largest systematic dependence in our results
comes from the functional form of $\rho(u)$. Various simple polynomials have been used,
as described in Table~\ref{tab:he_result}. As the reader can check, all
fit ans\"atze produce results for $\mbar(\mu_0/2)/\mbar(\mu_{\rm had})$
that agree within one sigma. As final result we choose the fit with
$n_r=3$ and $\rho(u)=\rho_0 + \rho_1u+\rho_2u^2$, which has the
best $\chi^2/{\rm dof}$, and yields
\begin{equation}
  \mbar(\mu_0/2)/\mbar(\mu_{\rm had}) = 0.5226(43)\,.
\end{equation}
The corresponding $\tau(x)$ function is obtained from Eq.~\eqref{eq:taugf}
by using the coefficients
\begin{equation}
f_0 = 1.28493\,, ~ f_1=-0.292465\,, ~ f_2=0.0606401\,, ~ f_3=-0.00291921
\end{equation}
together with the known coefficients for the $\beta$ function
in Eq.~\eqref{eq:betagf}. The covariance between the $f_n$ parameters reads
\begin{equation}
  \begin{split}
    {\rm cov}(f_i,f_j) = 
    &\left(
      {\footnotesize
      \begin{array}{cccc}
  \phantom{+}2.33798\times 10^{-2} & -1.47011\times 10^{-2} &  \phantom{+}2.81966\times
                                                   10^{-3} &
                                                             -1.66404\times
                                                             10^{-4} \\
 -1.47011\times 10^{-2} &  \phantom{+}9.54563\times 10^{-3} & -1.87752\times
                                                   10^{-3} &
                                                             \phantom{+}1.12962\times
                                                           10^{-4} \\
  \phantom{+}2.81966\times 10^{-3} & -1.87752\times 10^{-3} &  \phantom{+}3.78680\times
                                                   10^{-4} &
                                                             -2.32927\times
                                                             10^{-5} \\
 -1.66404\times 10^{-4} &  \phantom{+}1.12962\times 10^{-4} & -2.32927\times
                                                   10^{-5} &
                                                             \phantom{+}1.46553\times
                                                           10^{-6} \\
    \end{array}
  }
  \right)\,.
  \end{split}
\end{equation}
Finally, the covariance between the $p_i$ and the $f_j$ parameters is given by
\begin{equation}
  \begin{split}
    {\rm cov}(f_i,p_j) = 
    &\left(
      {\footnotesize
      \begin{array}{ccc}
 -3.83814\times 10^{-2} &  \phantom{+}1.25288\times 10^{-2} &
                                                              -8.87274\times
                                                              10^{-4} \\
  \phantom{+}1.58985\times 10^{-2} & -5.23410\times 10^{-3} &
                                                              \phantom{+}3.68003\times
                                                            10^{-4} \\
 -1.85736\times 10^{-3} &  \phantom{+}5.57473\times 10^{-4} &
                                                              -3.58668\times
                                                              10^{-5} \\
  \phantom{+}6.39488\times 10^{-5} & -1.45530\times 10^{-5} &
                                                              \phantom{+}5.96571\times
                                                             10^{-7} \\
    \end{array}
  }
  \right)\,.
  \end{split}
\end{equation}
The functional form of the anomalous dimension with its uncertainty
can thus be easily reproduced, and is shown in Fig.~\ref{fig:tau}.
Together with the result for $\tau$ in the high-energy region discussed
in Section~\ref{sec:sf}, and the result in Eq.~(\ref{eq:MovmSF}), one has
then all the ingredients needed to reconstruct the scale dependence
of light quark masses in the full energy range relevant to SM physics,
as shown in Fig.~\ref{fig:runningmass}.

Let us end this section by pointing out that the coefficient $f_0=1.2(2)$
is compatible within $1.5\sigma$ with the one-loop perturbative prediction $d_0/b_0$. This
is quite surprising, especially taking into account that the
$\beta$ function is not compatible with the one-loop functional
form with coefficient $b_0$. This in turn means that also the
anomalous dimension $\tau$ is poorly approximated by one-loop perturbation
theory. One is thus driven to conclude that
the agreement of $f_0$ with LO perturbation theory is only apparent.

\section{Hadronic matching and total renormalisation factor}
\label{sec:had}

The last step in our strategy requires the computation of the PCAC quark mass renormalisation
factors $\Zm$ at hadronic scales, cf. Eq.~(\ref{eq:Zhad}).
The latter can be written as
\begin{gather}
\Zm(g_0^2,a\muhad) = \frac{\ZA(g_0^2)}{\ZP(g_0^2,a\muhad)}\,
\end{gather}
Since the values of the axial current normalisation $\ZA$ are
available from a 
separate computation~\cite{Bulava:2016ktf,Brida:2016rmy,ZAchiSF} (in
particular we use the precise values obtained thanks to the chirally
rotated SF setup~\cite{Sint:2010eh,Sint:2010xy,Brida:2014zwa}), in
order to obtain $\Zm$ we just need to determine $\ZP(g_0^2,a\muhad)$
at a fixed value of $\muhad$ for changing bare coupling $g_0^2$.
The values of $g_0^2$ have to be in the range used in large-volume simulations;
for practical purposes, we will thus target the interval $\beta = 6/g_0^2 \in[3.40,3.85]$
currently covered by CLS ensembles~\cite{Bruno:2014jqa,Mohler:2017wnb}.

Our strategy proceeds as follows. We first choose a value of
$\uhad=\gbar_{\rm\scriptscriptstyle GF}^2(\muhad)$ such that the
relevant $g_0^2$ range is covered using accessible
values of $L/a$. The precise value of $\uhad$ is fixed by simulating
at one of the finest lattices. Finally, other lattice sizes are simulated,
such that we can obtain an interpolating formula for $\ZP(g_0^2,a\muhad)$
as a function of $g_0^2$ along the line of constant physics fixed by $\uhad$.
We have set $\uhad=9.25$, fixed by simulating on an $L/a=20$ lattice at $\beta=3.79$.
Lattice sizes $L/a=16,12,10$ have then been used at smaller $\beta$, and two $L/a=24$
lattices have been added so that the finest $\beta=3.85$ point can be safely interpolated.
Using the results in~\cite{Bruno:2017gxd}, this value of $\uhad$ corresponds to an
energy scale
$\muhad = 233(8)~\MeV$.

Our simulation results are summarised in Table~\ref{tab:hadmatch}.
Deviations from the target values of $\uhad$ induce a small but visible effect
on $\ZP$. The measured values of the PCAC mass are often
beyond our tolerance $|Lm| \lesssim 0.001$ (cf. App.~\ref{app:sys}),
especially for the smaller lattice sizes. This is a consequence of the fact
that the interpolating formula for $\hopc$ as a function of $g_0^2$ loses
precision for $L/a=10$, and of the small cutoff effect induced by computing
at zero topological charge (which is part of our renormalisation condition).
Note that the values of $\ZP$ that enter the fit function
are never further away than two standard deviations from the
value on the lattice that defines the line of constant physics.
The fitted dependences on $g_0^2$, $\uGF$, and $Lm$ are thus very mild.

\begin{table}
\centering
\small
\begin{tabular}{CCCRLCC}\toprule
   L/a & \beta & \kappa & \multicolumn{1}{c}{$\uGF$} & \multicolumn{1}{c}{$Lm$} & \ZP(g_0^2,L/a) & \text{\#~cfg}\\
\midrule\\[-2.5ex]
 10 & 3.4000 & 0.1368040500 & 9.282(39) & -0.0221(31) & 0.3484(11) & 2489 \\ 
 10 & 3.4110 & 0.1367650000 & 9.290(32) & +0.0189(23) & 0.3526(10) & 4624 \\ 
 12 & 3.4800 & 0.1370389800 & 9.406(41) & -0.0115(32) & 0.3417(14) & 1828 \\ 
 12 & 3.4880 & 0.1370210000 & 9.393(43) & +0.0035(23) & 0.3430(15) & 2667 \\ 
 12 & 3.4970 & 0.1370629900 & 9.118(54) & -0.0102(32) & 0.3487(18) & 1491 \\ 
 16 & 3.6490 & 0.1371576500 & 9.423(39) & -0.0024(17) & 0.3430(19) & 4560 \\ 
 16 & 3.6576 & 0.1371541300 & 9.186(50) & -0.0039(18) & 0.3492(17) & 3079 \\ 
 16 & 3.6710 & 0.1371475600 & 9.045(91) & +0.0009(26) & 0.3526(28) & 1553 \\ 
 20 & 3.7900 & 0.1370480000 & 9.251(54) & -0.0008(11) & 0.3508(22) & 4133 \\ 
 24 & 3.8934 & 0.1368944446 & 9.382(56) & -0.0001(11) & 0.3474(20) & 4709 \\ 
 24 & 3.9122 & 0.1368621644 & 9.132(51) & +0.0001(7) & 0.3543(22) & 5086 \\ 
\bottomrule
\end{tabular}
\caption{Results for $\ZP$ in the $\LEs$ scheme, used to determine quark mass
         renormalization constants at $\uhad=9.25$.  Alongside the values of
         $\uGF$ and $\ZP$, we also quote the value of the PCAC mass $m$ in
         units of the physical lattice length, and the statistics for each
         ensemble.
}
\label{tab:hadmatch}
\end{table}

\begin{figure}[t!]
\begin{center}
\includegraphics[width=0.85\linewidth]{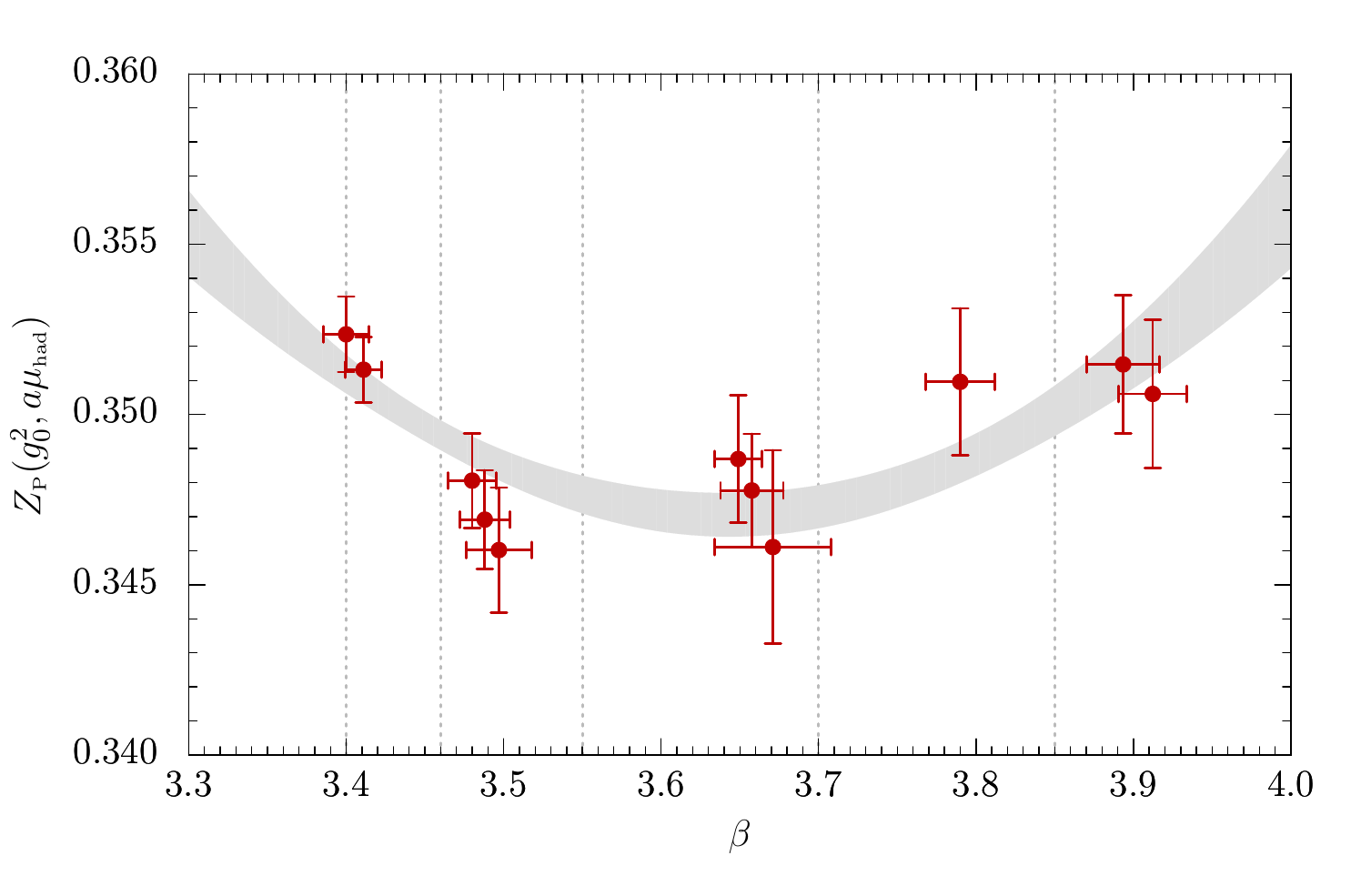}
\end{center}
\vspace{-5mm}
\caption{
Projection on the $(\beta,\ZP)$
plane of the fit (grey band) to the results for $\ZP$ at the hadronic matching point
$\uhad=9.25$ (filled red points).
The data points have been shifted to the target value of $\uGF$ and $Lm=0$ using the fit function.
Horizontal error bars have been assigned to reflect the uncertainties coming
from the value of $\uGF$ for each point, by defining $\Delta g_0^2/g_0^2 = \Delta\uGF/\uGF$.
Vertical dashed lines correspond to the $\beta$ values of CLS ensembles.
}
\label{fig:hadmatch}
\end{figure}

The measured values of $\ZP$ are fitted to a function of the form
\begin{gather}
\begin{split}
\ZP(g_0^2,\uGF,Lm) & = \ZP^{\rm\scriptscriptstyle had}(g_0^2)
+ t_{10}\,(\uGF-\uhad) + t_{20}\,(\uGF-\uhad)^2\\
& \qquad \qquad \quad
+ t_{01}\,Lm + t_{11}\,(\uGF-\uhad)Lm\,,\\
\ZP^{\rm\scriptscriptstyle had}(g_0^2) & = z_0 + z_1(\beta-\beta_0) + z_2(\beta-\beta_0)^2\,,
\end{split}
\end{gather}
where $\uhad=9.25$ and $\beta_0=3.79$. The terms with coefficients
$t_{ij}$ parameterise the small deviations from the intended
line of constant physics described above, while
$\ZP^{\rm\scriptscriptstyle had}$ is the interpolating function we are interested in.
Note that the ensembles are fully uncorrelated among them, but within each ensemble
the values of $\uGF$, $Lm$, and $\ZP$ are correlated, and we take this into
account in the fit procedure.
We have performed fits setting to zero various subsets of $t_{ij}$
coefficients; we quote as our preferred result the one with $t_{20}=t_{11}=0$,
for which $\chi^2/{\rm dof}=4.43/6$, and the coefficients for the interpolating
function $\ZP^{\rm\scriptscriptstyle had}$ read
\begin{subequations}
\label{eq:zpfitall}
\begin{equation}
\label{eq:zpfit}
z_0 = 0.348629 \,, \quad z_1 = 0.020921 \,, \quad z_2 = 0.070613 \,,
\end{equation}
with covariance matrix
\begin{equation}
\label{eq:zpfitcov}
{\rm cov}(z_i,z_j) = \left(\ba{rrr}
 0.375369 \times 10^{-6} &  0.429197 \times 10^{-6} & -0.186896 \times 10^{-5} \\
 0.429197 \times 10^{-6} &  0.268393 \times 10^{-4} &  0.686776 \times 10^{-4} \\
-0.186896 \times 10^{-5} &  0.686776 \times 10^{-4} &  0.212386 \times 10^{-3} \\
\ea\right)\,.
\end{equation}
\end{subequations}
The typical precision of the values of $\ZP$ extracted from the interpolating
function is thus at the few permille level.
The fit is illustrated in Fig.~\ref{fig:hadmatch}.

Now we can finally assemble the various factors entering Eqs.~(\ref{eq:factors},\ref{eq:Zhad}).
Using the results in Sec.~\ref{sec:gf} we can compute the running $\mbar(\muswi/2)/\mbar(\muhad)$
between the scheme-switching scale and hadronic matching scale,
and multiply it by the value of $M/\mbar(\muswi/2)$
given in Eq.~(\ref{eq:MovmSF}). Combining the errors in quadrature, we obtain
\begin{gather}
\label{eq:total_running}
\frac{M}{\mbar(\muhad)} = 0.9148(88)\,,
\end{gather}
which has a $0.96$\% precision; recall $\muhad = 233(8)~\MeV$.
It is important to stress that \req{eq:total_running} holds in the continuum,
i.e., it is independent of any detail of the lattice computation.
We can then take our interpolating formula for $\ZP$
and the known results for $\ZA$,
and build the total factor $\ZM$ that relates RGI masses to bare PCAC masses computed with
a non-perturbatively $\Oa$-improved fermion action and a tree-level Symanzik improved
gauge action,
\begin{gather}
\label{eq:ZRGI_def_w}
\ZM(g_0^2) = \frac{M}{\mbar(\muhad)}\frac{\ZA(g_0^2)}{\ZP(g_0^2,a\muhad)}\,.
\end{gather}
Recall that the dependence of $\ZM$ on $\muhad$ has to cancel exactly, up to
residual terms contributing to the cutoff effects of $\ZM(g_0^2)m(g_0^2)$.
Using as input the values of $\ZA$ from the chirally
rotated SF setup~\cite{Sint:2010eh,Sint:2010xy,Brida:2014zwa},
we quote the interpolating function
\begin{subequations}
\label{eq:ZRGI_wall}
\begin{equation}
\label{eq:ZRGI_w}
\begin{gathered}
\ZM(g_0^2) =   \frac{M}{\mbar(\muhad)}\times\left\{\ZM^{(0)} + \ZM^{(1)}(\beta-3.79) + \ZM^{(2)}(\beta-3.79)^2\right\}\,;\\[1.0ex]
\ZM^{(0)} =  2.270073\,, \quad
\ZM^{(1)} =  0.121644\,, \quad
\ZM^{(2)} = -0.464575\,,
\end{gathered}
\end{equation}
with covariance matrix
\begin{equation}
\label{eq:ZRGI_wcov}
{\rm cov}(\ZM^{(i)},\ZM^{(j)}) = \left(\ba{rrr}
 0.164635 \times 10^{-4} & 0.215658 \times 10^{-4} & -0.754203 \times 10^{-4} \\
 0.215658 \times 10^{-4} & 0.121072 \times 10^{-2} &  0.308890 \times 10^{-2} \\
-0.754203 \times 10^{-4} & 0.308890 \times 10^{-2} &  0.953843 \times 10^{-2} \\
\ea\right)\,.
\end{equation}
\end{subequations}
The quoted errors only contain the uncertainties from the determination of
$\ZA$ and $\ZP$ at the hadronic scale.
As remarked in~\cite{Capitani:1998mq},
the error of the total running factor $M/\mbar(\muhad)$ in Eq.~(\ref{eq:total_running}) has
to be added in quadrature to the error in the final result for the RGI mass, since it
only affects the continuum limit, and should not be included in the extrapolation
of data for current quark masses to the continuum limit.

Finally, we note that our results can be also used to obtain renormalised quark masses when a
twisted-mass QCD Wilson fermion regularisation~\cite{Frezzotti:2000nk} is employed in the computation.
In that case the bare PCAC mass coincides with the bare twisted mass parameter,
which can be renormalised with
\begin{gather}
\Zm^{\rm\scriptscriptstyle tm}(g_0^2,a\muhad) = \frac{1}{\ZP(g_0^2,a\muhad)}\,.
\end{gather}
The total renormalisation factor then acquires the form
\begin{gather}
\label{eq:ZRGI_def_tmqcd}
\ZM^{\rm\scriptscriptstyle tm}(g_0^2) = \frac{M}{\mbar(\muhad)}\frac{1}{\ZP(g_0^2,a\muhad)}\,,
\end{gather}
and values can be obtained by directly using our interpolating
formula for $\ZP^{\rm\scriptscriptstyle had}$.
The same comment about the combination of uncertainties as above applies.
The values of $\ZM$ and $\ZM^{\rm\scriptscriptstyle tm}$
at the $\beta$ values of CLS ensembles are provided in Table~\ref{tab:ZRGI}.

Eqs.~(\ref{eq:total_running},\ref{eq:zpfitall},\ref{eq:ZRGI_wall}) are the final, and most important, results of this work.
We stress once more that the result for $M/\mbar(\muhad)$, which is
by far the most computationally demanding one, holds
in the continuum, and is independent of the lattice regularisation
employed in its determination, as well as any other computational detail.
The expressions for $\ZM$ and $\ZM^{\rm\scriptscriptstyle tm}$,
on the other hand, depend on the action used in the computation, and hold for
a non-perturbatively $\Oa$-improved fermion action and a tree-level Symanzik improved gauge action.
Repeating the computation for a different lattice action would only
require obtaining the values of $\ZA$ and $\ZP$ in the appropriate
interval of values of $\beta$, at small computational cost.

\begin{table}
\centering
\begin{tabular}{ccc}
\toprule
$\beta$ & $\ZM$ & $\ZM^{\rm\scriptscriptstyle tm}$ \\
\midrule
$3.40$ & 1.9684(35) & 2.6047(42) \\
$3.46$ & 1.9935(27) & 2.6181(33) \\
$3.55$ & 2.0253(33) & 2.6312(42) \\
$3.70$ & 2.0630(38) & 2.6339(48) \\
$3.85$ & 2.0814(45) & 2.6127(55) \\
\bottomrule
\end{tabular}
\caption{Values of the renormalisation factors $\ZM$
         and $\ZM^{\rm\scriptscriptstyle tm}$ connecting RGI and bare PCAC
         masses in the standard Wilson and twisted-mass cases, respectively, at
         the values of $\beta$ for CLS ensembles. Recall that the quoted errors
         do not contain the contribution from the running factor $M/\mbar(\muhad)$.
         The correlations between the errors at different $\beta$ can be
         readily obtained from the covariance matrices provided in
         Eqs.~(\ref{eq:ZRGI_wcov}) and~(\ref{eq:zpfitcov}), respectively.
}
\label{tab:ZRGI}
\end{table}

\section{Conclusions}
\label{sec:con}

In this paper we have performed a fully non-perturbative, high-precision determination
of the quark mass anomalous dimension in $\NF=3$ QCD, spanning from the electroweak
scale to typical hadronic energies. Alongside the companion non-perturbative determination 
of the $\beta$ function in~\cite{Brida:2016flw,DallaBrida:2016kgh}, this completes the first-principles determination
of the RG functions of fundamental parameters for light hadron physics.
Together with the determination of the $\lQCD$ parameter~\cite{Bruno:2017gxd}
and the forthcoming publication of renormalised quark masses~\cite{ALPHAmass},
a full renormalisation programme of $\NF=3$ QCD will have been achieved.
For the purpose of the latter computation, we have also provided a precise
computation of the matching factors required to obtain renormalised quark masses 
from PCAC bare quark masses obtained from simulations based on CLS $\NF=2+1$
ensembles~\cite{Bruno:2016plf,Herdoiza:2017bcc}.
The total uncertainty introduced by the matching factor to RGI quark
masses is at the level of $1.1\%$.

A slight increase in this precision is achievable within the same framework.
This would require however a significantly larger numerical effort, adding larger
lattices and hence finer lattice spacings to the continuum limit extrapolation,
and augmenting the precision for the tuning of bare parameters.
One lesson from the present work is that
the $1\%$ ballpark is not much above the irreducible systematic uncertainty
achievable with the methodology employed. Improvements of the latter will thus be
necessary to reduce the uncertainty purely due to renormalisation to the few permille level.
It is important to stress that the uncertainty is completely dominated by the
contribution from the non-perturbative running at low energies; another lesson
from the present work is that, at this level of precision, use of perturbation
theory in the few-GeV region is not necessarily satisfactory.

Apart from increasing the precision of the computation, one obvious next step
is the inclusion of heavier flavours. This would ideally result in reaching
sub-percent renormalisation-related uncertainties in first-principles computations of the
charm and bottom quark masses, whose values play a key role in frontier studies
of B- and Higgs physics.
First steps in this direction, at the level of the computation of the running
coupling, have already been
taken~\cite{Knechtli:2017xgy,Cali:2017brl,Knechtli:2017pxe}.

\section*{Acknowledgements}

P.F., C.P. and D.P. acknowledge support through the Spanish MINECO
project FPA2015-68541-P and the Centro de Excelencia Severo Ochoa
Programme SEV-2012-0249 and SEV-2016-0597. C.P. and D.P. acknowledge the kind
hospitality offered by CERN-TH at various stages of this work.
The simulations reported here were performed on  the following HPC
systems: Altamira, provided by IFCA at the University of Cantabria, on
the FinisTerrae-II machine provided by CESGA (Galicia Supercomputing
Centre), and on the Galileo HPC system provided by CINECA. FinisTerrae
II was funded by the Xunta de Galicia and the Spanish MINECO under the
2007-2013 Spanish ERDF Programme.
Part of the simulations reported in \res{sec:had} were performed on a dedicated HPC cluster at CERN.
We thankfully acknowledge the
computer resources offered and the technical support provided by the
staff of these computing centers.
We are grateful to J.~Koponen for helping us complete some of
the simulations discussed in \res{sec:had} in the
concluding stages of this project.
We are indebted to our fellow members of the ALPHA Collaboration for
many valuable discussions and the synergies developed with other
renormalisation-related projects; special thanks go
to M.~Dalla~Brida, T.~Korzec, R.~Sommer, and S.~Sint.

\cleardoublepage
\begin{appendix}
\section{Systematic uncertainties in the determination of step scaling functions}
\label{app:sys}

\subsection{Tuning of the critical mass}

The tuning of the chiral point, as part of setting our lines of constant physics,
is treated in detail in~\cite{Nf3tuning}. Briefly, a set of tuning runs at various values
of $\beta$ and $\kappa$ are used to compute the PCAC mass $m$ in~Eq.~(\ref{eq:pcac}),
and interpolate at fixed $\beta$ values for $\hopc$ such that $|Lm| \leq 0.001$,
with an uncertainty of at most the same order. This implies that the values
of the quark mass at which the renormalisation condition in Eq.~(\ref{eq:ZP})
is imposed are not exactly zero.

\begin{figure}[t!]
\begin{center}
\includegraphics[width=0.7\linewidth]{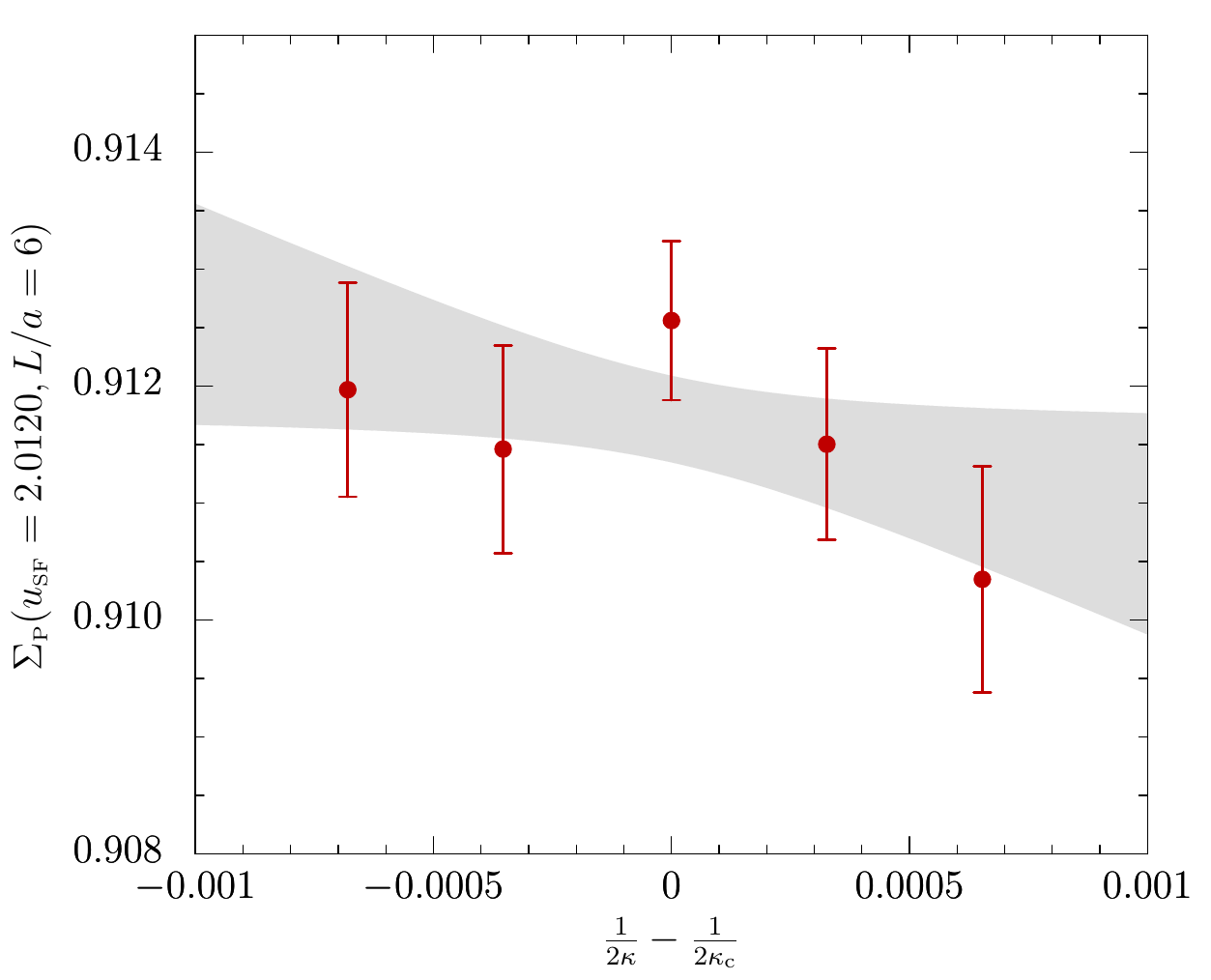}
\end{center}
\vspace{-5mm}
\caption{
Variation of the SSF $\SigmaP(\uSF=2.0120,L/a=6)$ with the value of the quark mass.
The shadowed band is a linear fit to the data.
}
\label{fig:kcsys}
\end{figure}

In order to assess the relevant systematics, we have performed dedicated
simulations at the strongest coupling covered in the $\HEs$ scheme, $\uSF=2.0120$,
for which $\SigmaP(2.0120,L/a=6)$ has been computed at the value
of $\beta=6.2735$ indicated in Table~\ref{tab:ZP_SF}, and four different values of $\kappa$,
besides the nominal one for $\hopc$ given in Table~\ref{tab:ZP_SF}. This is
expected to be the simulation point within the high-energy regime where
systematics may have a stronger impact. The result of this exercise is shown
in Fig.~\ref{fig:kcsys}. A linear fit to the data allows to estimate the
slope coefficient
\begin{gather}
\rho_{\hopc} \equiv \frac{1}{L}\,\left.\frac{\partial\SigmaP}{\partial m}\right|_{u,L}\,,
\end{gather}
which can then be used to assign a systematic uncertainty to $\SigmaP$ as
\begin{gather}
\delta_{\hopc}\SigmaP = |\rho_{\hopc}|{\rm tol}(Lm)\,,
\end{gather}
where ${\rm tol}(Lm)=0.001$ is our tolerance for defining the critical point.
We obtain $\rho_{\hopc}=-0.15(15)$, which yields for the systematic uncertainty
$\delta_{\hopc}\SigmaP = 0.00015(15)$.\footnote{Our value for
the slope $\rho_{\hopc}$ is in the same ballpark as the ones obtained in~\cite{kcnf2},
where a similar study using $\NF=2$ simulations was performed at values of the SF
coupling $\uSF=0.9793$ and $\uSF=2.4792$, finding $\rho_{\hopc}=-0.0755(10)$
and $\rho_{\hopc}=-0.1130(27)$, respectively.}

Besides being compatible
with zero within $1\sigma$, the central value is more than four times smaller than the
statistical uncertainty quoted for $\SigmaP(2.0120,L/a=6)$ in Table~\ref{tab:ZP_SF}. For larger lattices
or smaller renormalised couplings these systematics are expected to decrease further.
On the other hand, they will increase as the renormalised coupling increases,
but there is no obvious reason why its relative size with respect to the statistical uncertainty
will grow as well.
We thus conclude that the systematic uncertainty related to the tuning to zero quark mass
is negligible at the level of precision we attain in the computation of $\sigmaP$.

\subsection{Tuning of the gauge coupling}

Our lines of constant physics are formally defined by a nominal value of either
the SF or the GF coupling, that is to be kept fixed in the lattices at which
the denominator of Eq.~(\ref{eq:ssfcl}) is computed. In practice, this is so
only within some finite precision, due to two different reasons:
\begin{enumerate}
\item Both couplings $\uSF$ or $\uGF$ are computed with finite precision.
\item The $\HEs$ computations of the coupling and the corresponding tuning of $\beta$ and $\kappa$ are 
performed using independent ensembles with respect to the
ones employed for the computation of $\ZP$, for reasons explained in
\res{sec:str}. The resulting lines of constant physics are expected to differ by $\Oasq$ effects.
\end{enumerate}

In order to assess whether the finite precision on the value of the gauge coupling
has an impact on the computation of the continuum $\sigmaP$, we have:
\begin{enumerate}
\item Repeated the continuum-limit extrapolations of $\SigmaP$ at fixed $u$, introducing
a horizontal error on the value of $(a/L)^2$ propagated from the uncertainty on $u$
at each value of $a/L$. To that purpose one can use either the perturbative relation
between $u$ and $L$, or the known non-perturbative $\beta$ functions (cf.~\res{sec:str}),
with little practical differences.
\item Repeated fits to the continuum-extrapolated $\sigmaP$ as a function of $u$,
introducing an uncertainty on $u$ that covers the spread of the computed values
along that particular line of constant physics.
\end{enumerate}
In either case, we have found that the impact of introducing the additional uncertainties
in the final description of $\sigmaP$ are completely negligible within our current level of
precision. This source of systematic uncertainty is therefore ignored in our final analyses.

\subsection{Perturbative values of boundary improvement coefficients}

In our computation, we use perturbative values for the coefficients $\ict$ and $\icttil$
that appear in Schr\"odinger Functional boundary $\Oa$ improvement counterterms,
employing the highest available order for the relevant lattice action.
In computations in the high-energy region, where the plaquette gauge action is used,
we are able to use the corresponding two-loop value of $\ict$~\cite{Bode:1999sm}.
In the low-energy region, where the gauge action is tree-level Symanzik improved,
the two-loop coefficient is not known, and we take the one-loop value.
In the case of $\icttil$, we employ the one-loop value~\cite{Luscher:1996vw} throughout.
Contributions from these boundary counterterms to the step scaling function $\sigmaP$
start at two-loop order in perturbation theory~\cite{Sint:1998iq}, and the effects of perturbative
truncation in the values of $\ict,\icttil$ are therefore expected to be small.
A careful study of these systematic uncertainties in the \HEs~scheme was carried
out in the $\NF=0$~\cite{Capitani:1998mq} and $\NF=2$~\cite{DellaMorte:2005kg} computations.
Especially in the former case, a precise statement could be made that perturbative
truncation effects do not change the result for the continuum limit of $\SigmaP$
even at the largest values of the coupling. In our computation, we have performed
a dedicated analysis at two values of the coupling, to check the size of the
resulting effects in both the $\HEs$ and $\LEs$ schemes.

In order to have a quantitative handle on the effect from a shift on boundary improvement coefficients,
let us formally expand $\ZP$, considered as a function of, e.g., $\ict$,
in a power series of the form
\begin{gather}
\ZP + \frac{\partial\ZP}{\partial\ict}
\left(\frac{a}{L}\right)
\Delta\ict + \ldots
\end{gather}
where $\Delta\ict$ is the deviation with respect to the
value of $\ict$ at which $\ZP$ is computed.
The factor $(a/L)$ is made explicit to stress that the perturbative truncation
is leaving uncancelled $\Oa$ terms, which we are parameterising.
By simulating at a number of values of $\ict$, keeping all other
simulation parameters fixed, it is possible to estimate the slope
$(\partial\ZP/\partial\ict)$.
A systematic uncertainty can then be assigned to $\ZP$ as
\begin{gather}
\delta_{\ict}\ZP \approx
\left|\frac{\partial\ZP}{\partial\ict}\right|
\left(\frac{a}{L}\right)
\delta\ict\,,
\end{gather}
where $\delta\ict$ is some conservative
estimate of the perturbative truncation error. Linear error propagation
then yields the corresponding systematic uncertainty on $\SigmaP=\ZP(2L)/\ZP(L)$
as
\begin{gather}
\frac{\delta_{\ict}\SigmaP}{\SigmaP} \approx
\left|\frac{\partial\ZP}{\partial\ict}
\left[
\frac{1}{2\ZP(2L)}-\frac{1}{\ZP(L)}
\right]
\right|
\left(\frac{a}{L}\right)
\delta\ict\,.
\end{gather}
The systematic uncertainty due to the truncation in the value of $\icttil$
can be estimated in exactly the same way.

\begin{figure}[p]
\begin{center}
\begin{minipage}[t!]{0.48\textwidth}
\hspace*{-3mm}
\includegraphics[width=1.05\textwidth]{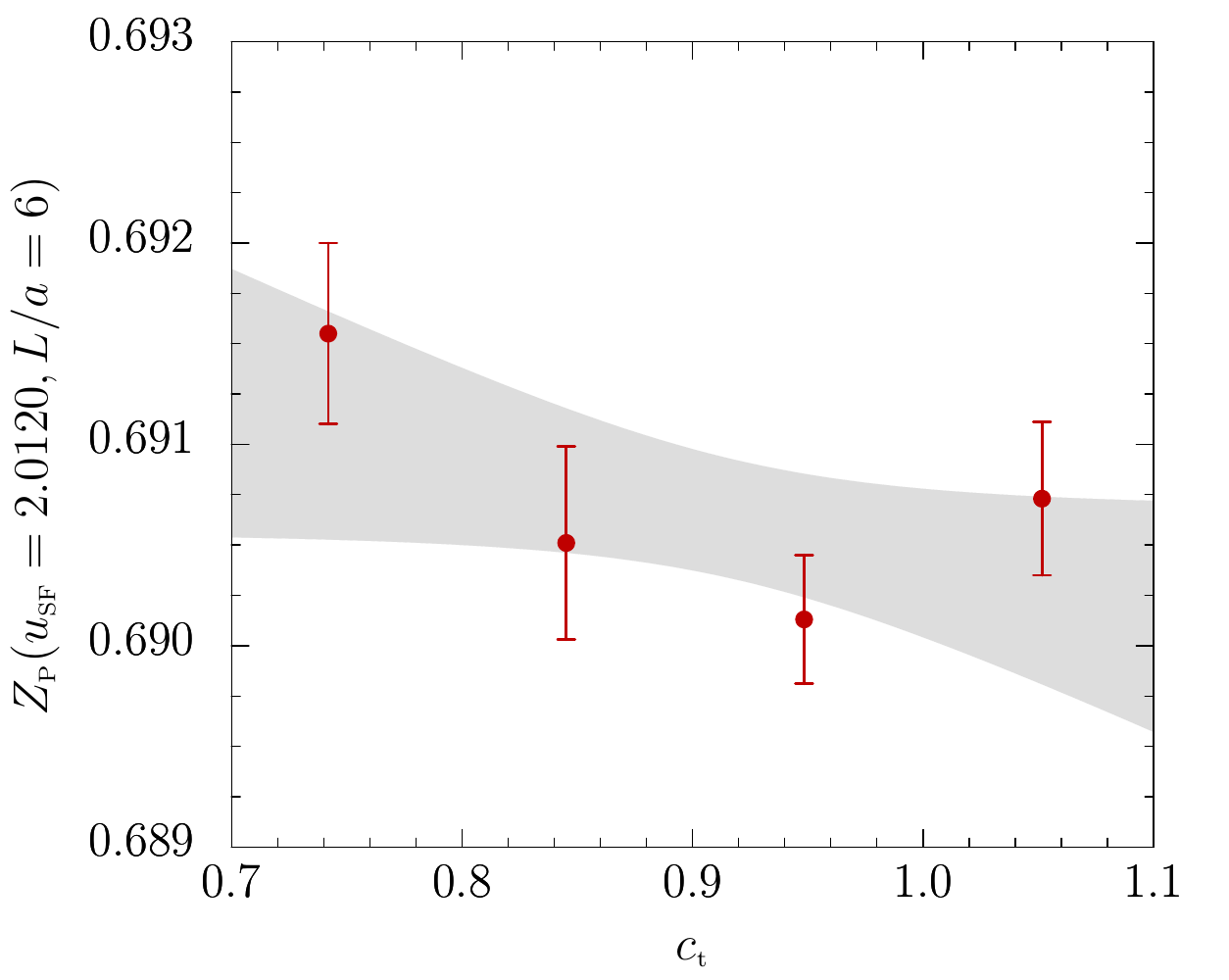}
\end{minipage}
\hspace{0mm}
\begin{minipage}[t!]{0.48\textwidth}
\hspace*{2mm}
\includegraphics[width=1.05\textwidth]{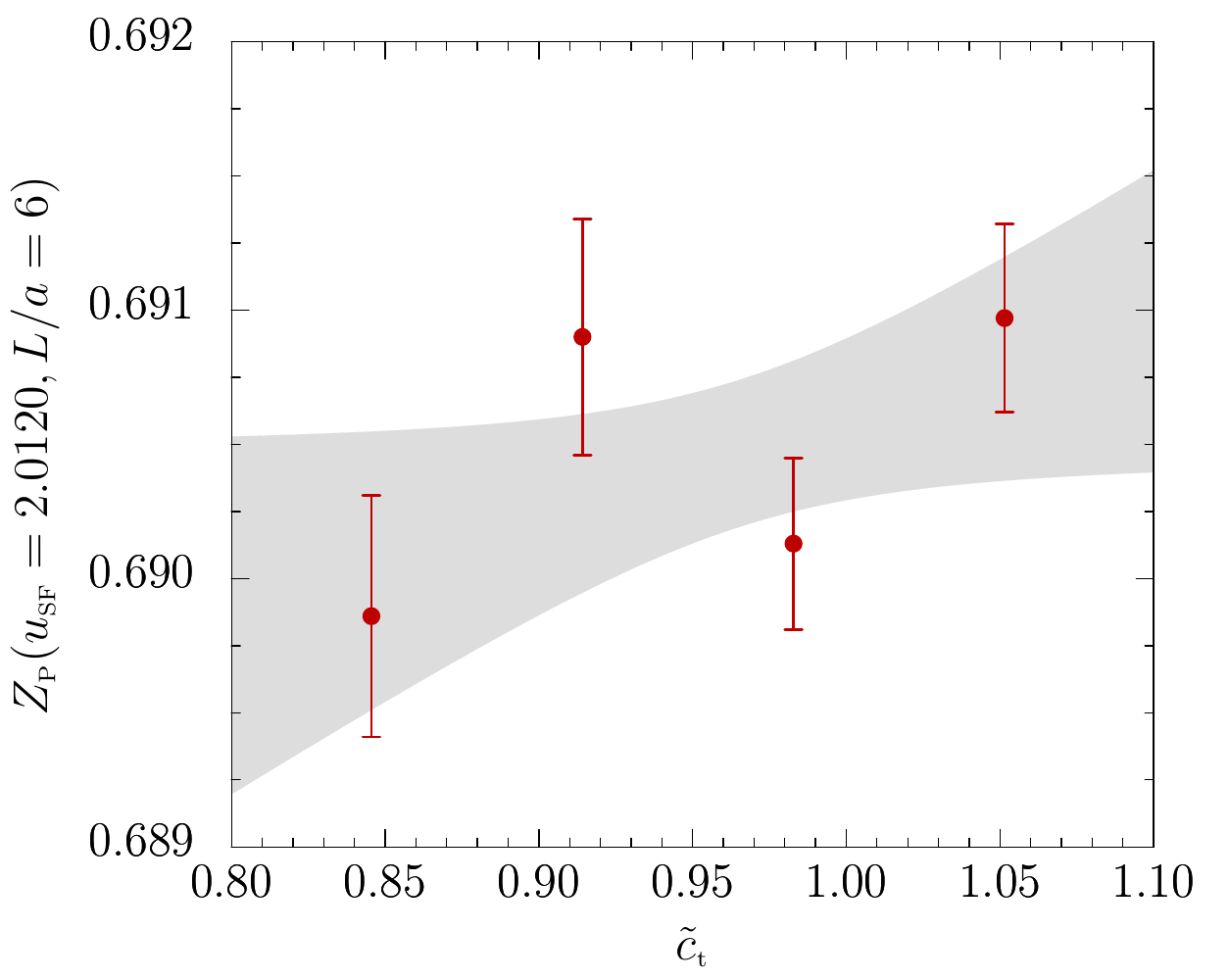}
\end{minipage}
\begin{minipage}[t!]{0.48\textwidth}
\hspace*{-3mm}
\includegraphics[width=1.05\textwidth]{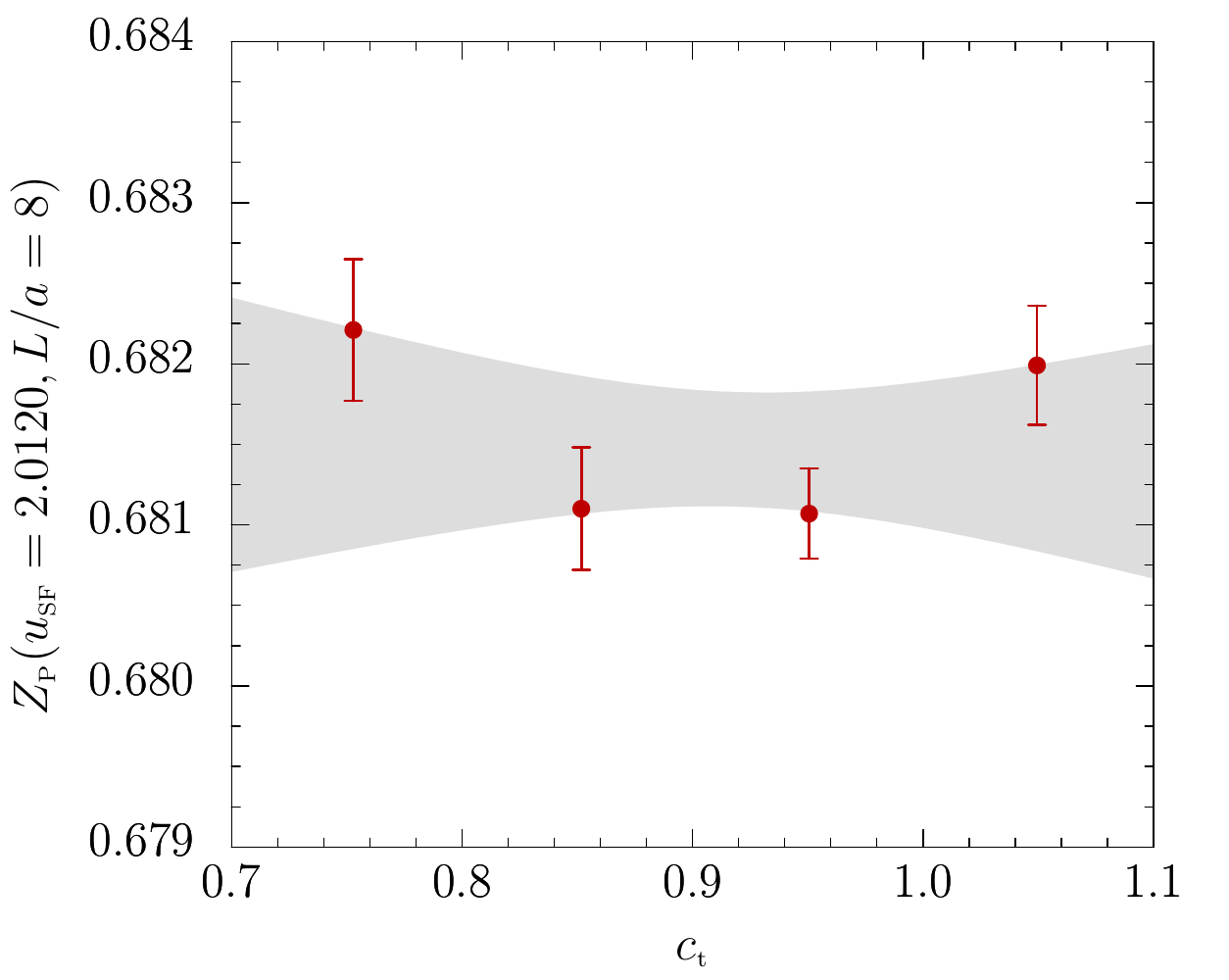}
\end{minipage}
\hspace{0mm}
\begin{minipage}[t!]{0.48\textwidth}
\hspace*{2mm}
\includegraphics[width=1.05\textwidth]{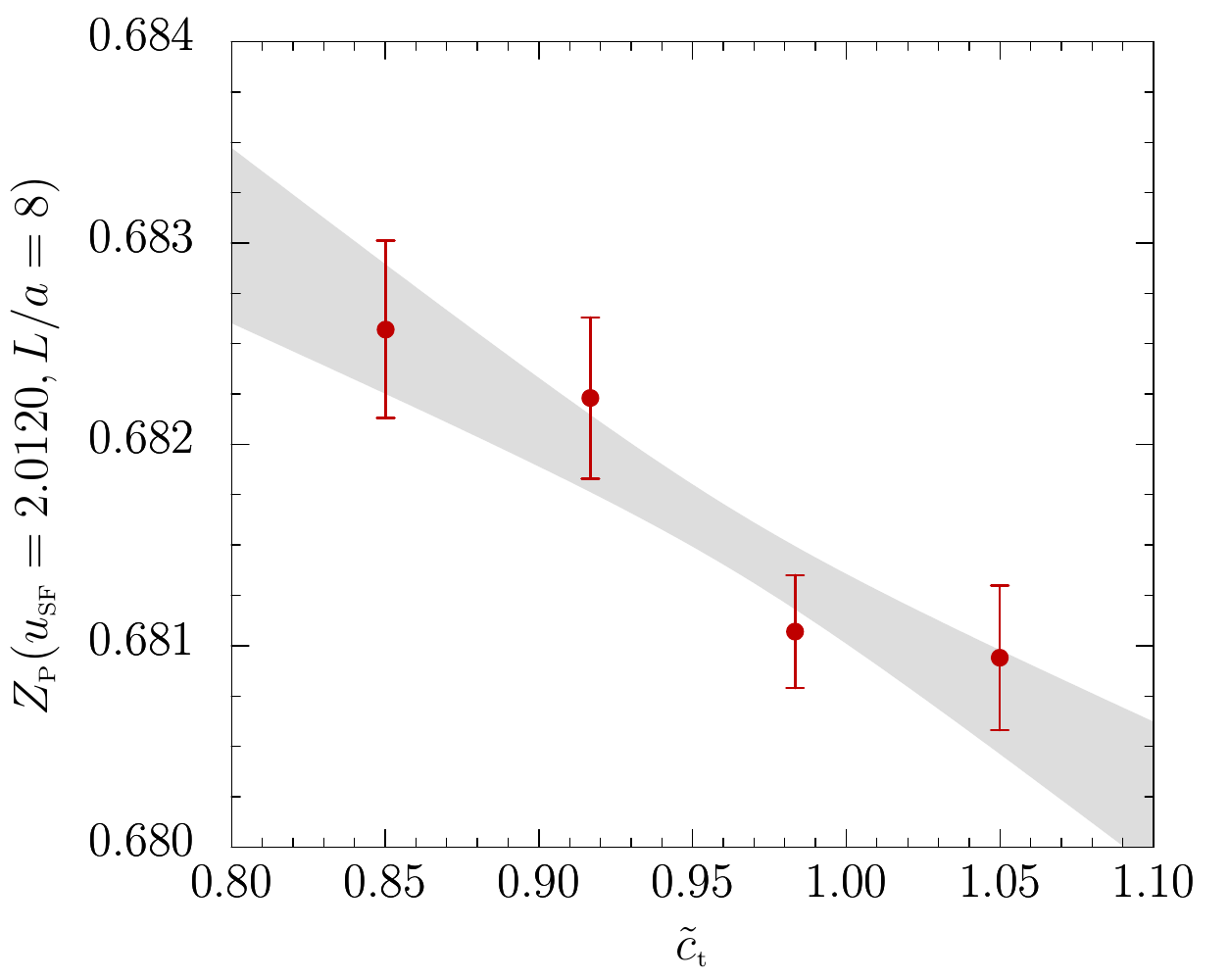}
\end{minipage}
\begin{minipage}[t!]{0.48\textwidth}
\hspace*{-3mm}
\includegraphics[width=1.05\textwidth]{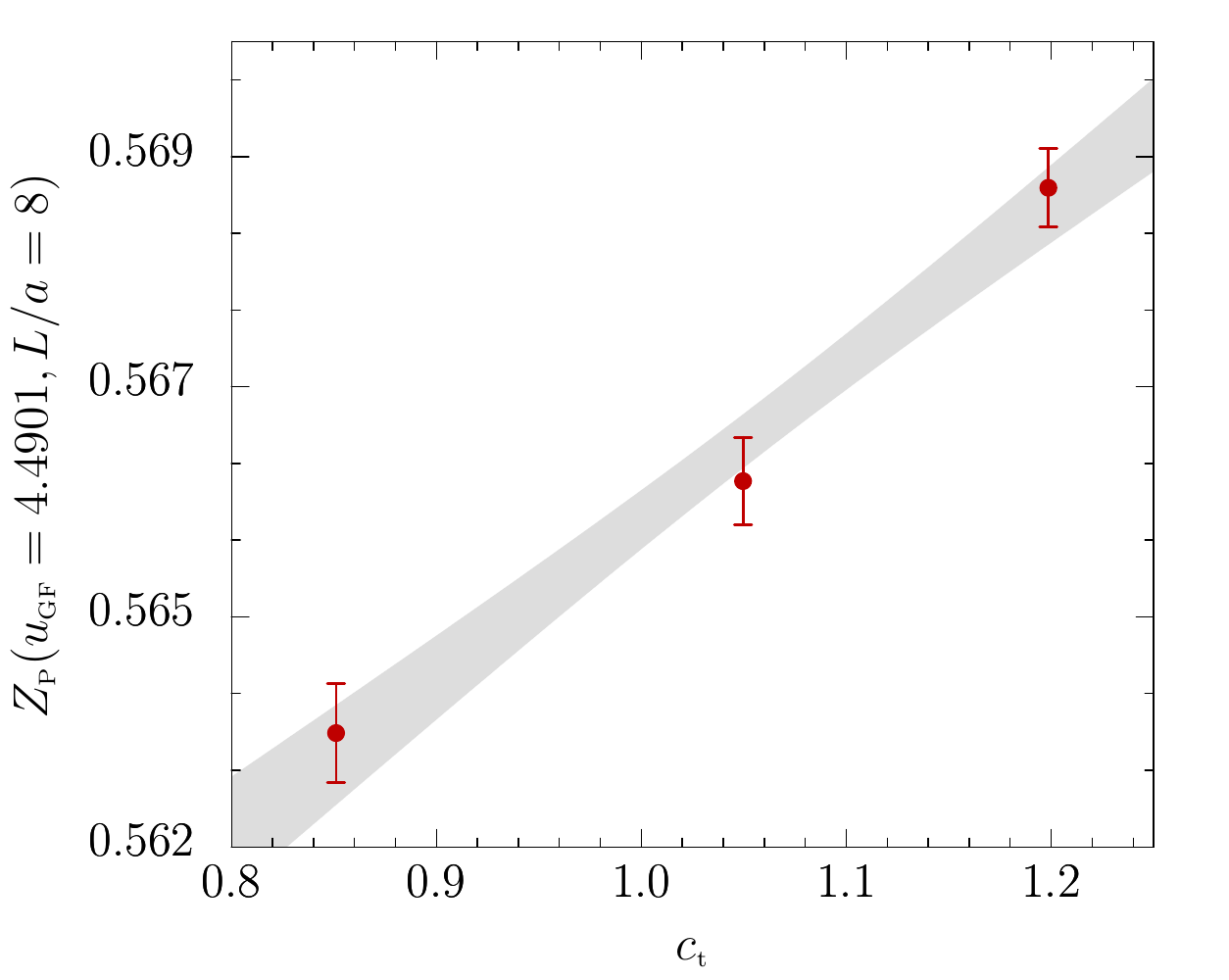}
\end{minipage}
\hspace{0mm}
\begin{minipage}[t!]{0.48\textwidth}
\hspace*{2mm}
\includegraphics[width=1.05\textwidth]{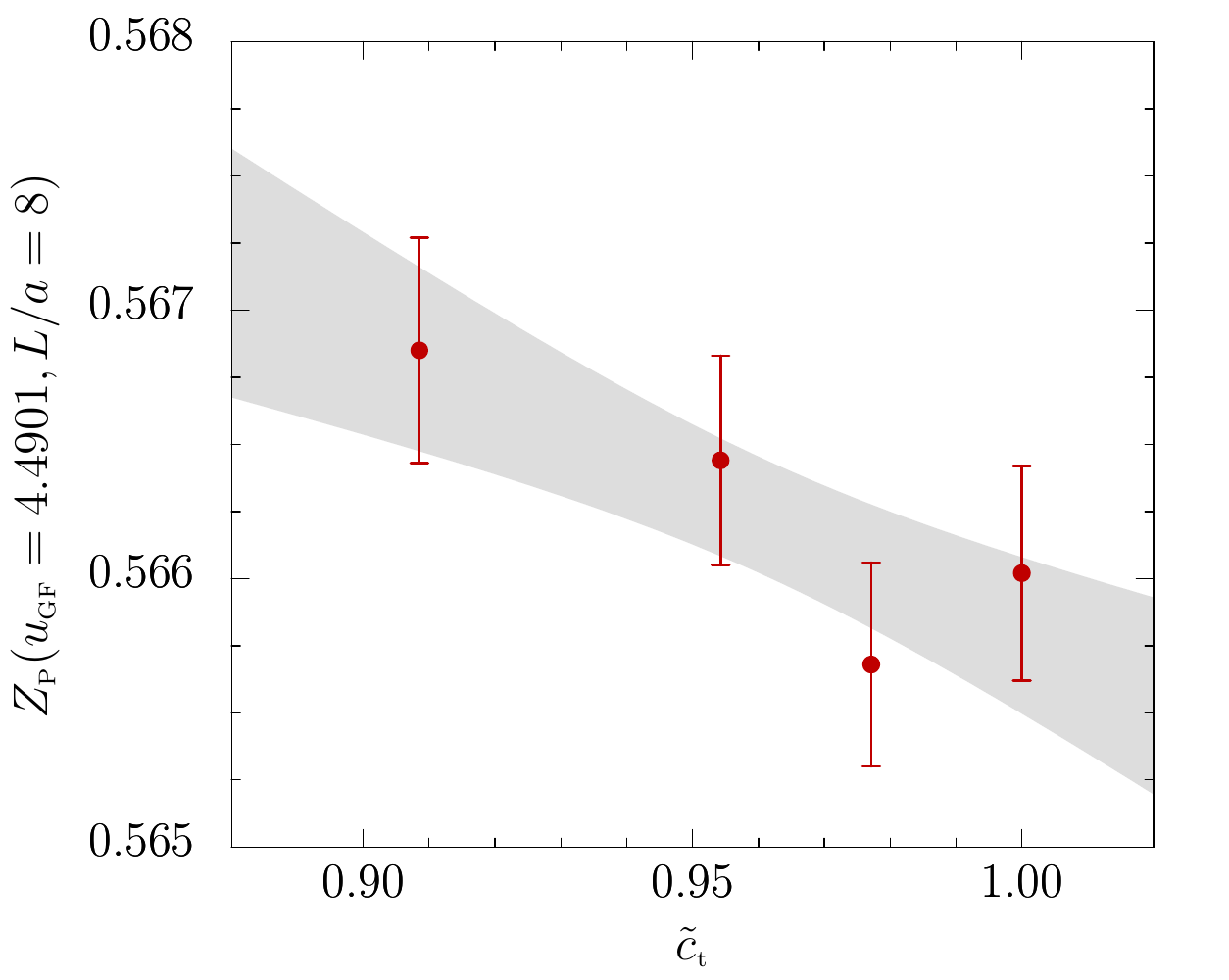}
\end{minipage}
\end{center}
\vspace{-5mm}
\caption{
Variation of $\ZP$ with the value of the boundary improvement
coefficients $\ict$ and $\icttil$. The shadowed bands are linear fits to the data.
}
\label{fig:bisys}
\end{figure}

We have performed dedicated simulations to estimate $(\partial\ZP/\partial\ict)$
and $(\partial\ZP/\partial\icttil)$ in the $\HEs$ scheme at $u=2.0120$
and in the $\LEs$ scheme at $u=4.4901$,
considering several values of $\ict$ and $\icttil$ in an interval given by artificially
augmenting the size of the perturbative correction to the tree-level value~$1$ 
by factors of up to $4$. The simulations are performed in $L/a=6$ and $L/a=8$ lattices,
which should be affected by the largest uncertainty.
The results are illustrated in Fig.~\ref{fig:bisys}.
By fitting the results for $\ZP$ linearly in the value of the improvement coefficient
we find
\begin{alignat}{4}
& \left.\frac{\partial\ZP}{\partial\ict}\right|_{\scriptscriptstyle\HEs;L/a=6} && = -0.016(15) \,, \qquad\qquad
&& \left.\frac{\partial\ZP}{\partial\icttil}\right|_{\scriptscriptstyle\HEs;L/a=6} && = 0.022(21) \,;\\
& \left.\frac{\partial\ZP}{\partial\ict}\right|_{\scriptscriptstyle\HEs;L/a=8} && = -0.003(26) \,, \qquad\qquad
&& \left.\frac{\partial\ZP}{\partial\icttil}\right|_{\scriptscriptstyle\HEs;L/a=8} && = -0.074(19) \,;\\
& \left.\frac{\partial\ZP}{\partial\ict}\right|_{\scriptscriptstyle\LEs;L/a=8} && =  0.110(14) \,, \qquad\qquad
&& \left.\frac{\partial\ZP}{\partial\icttil}\right|_{\scriptscriptstyle\LEs;L/a=8} && = -0.090(4) \,.
\end{alignat}
If now we use values of $\Delta\ict$ and $\Delta\icttil$ given by
$1-\ict^{\rm\scriptscriptstyle pert}$ and $1-\icttil^{\rm\scriptscriptstyle pert}$, respectively ---
i.e., we assign a $100\%$ uncertainty to the perturbative correction to the tree-level value~$1$ ---
we finally obtain
\begin{alignat}{7}
& \uSF && =2.0120\,,~L/a=6: \qquad
&& \delta_{\ict}\SigmaP   && \approx 0.00014 \,, \quad
&& \delta_{\icttil}\SigmaP && \approx 0.000063 \,;\\
& \uSF && =2.0120\,,~L/a=8: \qquad
&& \delta_{\ict}\SigmaP   && \approx 0.000019 \,, \quad
&& \delta_{\icttil}\SigmaP && \approx 0.00015 \,;\\
& \uGF && =4.4901\,,~L/a=8: \qquad
&& \delta_{\ict}\SigmaP   && \approx 0.00037 \,, \quad
&& \delta_{\icttil}\SigmaP && \approx 0.00014 \,.
\end{alignat}
In all cases, these figures are much smaller than the quoted
statistical error of $\SigmaP$. This justifies neglecting this source of systematic
uncertainty in our analysis.

\section{Simulation details}\label{app:sdetail}

Our data is partly based on ensembles that have been produced in conjunction
with the running coupling project~\cite{Brida:2016flw,DallaBrida:2016kgh}.  The
low-energy ensembles are common to both projects, such that we had to take into
account correlations between $\ZP$ and $\uGF$ as explained in
section~\ref{sec:gf}. To reach the desired precision in the computation of the
mass anomalous dimension, the statistics on some of those ensembles had to be
increased. Hence, we give a comprehensive list of the corresponding simulations
in Table~\ref{tab:GFruns} which enter our data analysis. 

The high-energy part is different in the respect that the computation of the SF
coupling $\uSF$, defining the lines of constant physics, is done with
non-vanishing background gauge field, while $\ZP$ is better computed with zero
background field. As a result we have produced an independent set of ensembles
summarised in Table~\ref{tab:SFruns}. The bare parameters are inherited from
the line of constant physics condition~\cite{Brida:2016flw,Nf3tuning}.

Tables~\ref{tab:SFruns},~\ref{tab:GFruns} list the line of constant physics
$(L/a,\beta,\kappa)$ for several fixed values of the renormalized coupling
$u=\gbar^2(L)$, corresponding to a fixed scale $L$ in the continuum. The data
are complemented by the number of measurements $\Nms$ and their separation $\tms$
in molecular-dynamics (MD) units, the measured integrated autocorrelation times
for $\ZP$, values of the dimensionless bare current quark mass $Lm_1$, and the
boundary-to-boundary correlator $f_1$.

\section{Tables}

\cleardoublepage
\begin{landscape}
\begin{table}[h]
  \centering
  \small
  \begin{tabular}{C@{\hspace{5mm}}CCC@{\hspace{5mm}}CCL@{\hspace{5mm}}CCC}\toprule
     \uSF & L/a & \beta   & \kappa     &\ZP(g_0^2,L/a)&\ZP(g_0^2,2L/a)&\SigmaPI(g_0^2,L/a) & \sigmaP(\uSF) & \rho_{\rm\scriptscriptstyle P}(\uSF) & \chi^2/{\rm dof} \\\midrule
          & 6   &  8.5403 & 0.13233610 & 0.80494(22) & 0.76879(24) &  0.95587(40)(78)  &                  &            &      \\
  1.11000 & 8   &  8.7325 & 0.13213380 & 0.79640(22) & 0.76163(34) &  0.95625(50)(10)  & 0.9577(10)       & -0.079(56) & 0.64 \\
          & 12  &  8.9950 & 0.13186210 & 0.78473(29) & 0.75167(59) &  0.95762(83)(24)  &                  &            &      \\  \midrule
          & 6   &  8.2170 & 0.13269030 & 0.79463(25) & 0.75594(25) &  0.95213(43)(82)  &                  &            &      \\
  1.18446 & 8   &  8.4044 & 0.13247670 & 0.78532(21) & 0.74943(39) &  0.95420(56)(10)  & 0.9558(10)       & -0.122(58) & 0.50 \\
          & 12  &  8.6769 & 0.13217153 & 0.77426(30) & 0.73937(49) &  0.95468(73)(25)  &                  &            &      \\  \midrule
          & 6   &  7.9091 & 0.13305720 & 0.78344(24) & 0.74256(25) &  0.94870(44)(88)  &                  &            &      \\
  1.26569 & 8   &  8.0929 & 0.13283120 & 0.77369(23) & 0.73483(43) &  0.94967(62)(11)  & 0.9512(10)       & -0.093(61) & 0.03 \\ 
          & 12  &  8.3730 & 0.13249231 & 0.76349(33) & 0.72601(49) &  0.95064(76)(27)  &                  &            &      \\  \midrule
          & 6   &  7.5909 & 0.13346930 & 0.77034(25) & 0.72691(26) &  0.94457(46)(94)  &                  &            &      \\
  1.3627  & 8   &  7.7723 & 0.13322830 & 0.76036(24) & 0.71893(42) &  0.94539(63)(11)  & 0.9482(11)       & -0.149(66) & 1.11 \\
          & 12  &  8.0578 & 0.13285365 & 0.74937(35) & 0.71036(52) &  0.94766(83)(29)  &                  &            &      \\  \midrule
          & 6   &  7.2618 & 0.13393370 & 0.75460(27) & 0.70808(31) &  0.9394(5)(10)    &                  &            &      \\
  1.4808  & 8   &  7.4424 & 0.13367450 & 0.74425(26) & 0.70004(33) &  0.94047(55)(13)  & 0.9415(11)       & -0.071(66) & 0.06 \\
          & 12  &  7.7299 & 0.13326353 & 0.73515(33) & 0.69193(42) &  0.94090(72)(31)  &                  &            &      \\  \midrule
          & 6   &  6.9433 & 0.13442200 & 0.73740(28) & 0.68692(29) &  0.9327(5)(11)    &                  &            &      \\
  1.6173  & 8   &  7.1254 & 0.13414180 & 0.72670(26) & 0.67983(49) &  0.93536(76)(14)  & 0.9359(15)       & -0.086(82) & 2.56 \\ 
          & 12  &  7.4107 & 0.13369922 & 0.71777(41) & 0.67071(75) &  0.9341(12)(3)    &                  &            &      \\  \midrule
          & 6   &  6.6050 & 0.13498290 & 0.71606(29) & 0.65960(32) &  0.9224(6)(12)    &                  &            &      \\
  1.7943  & 8   &  6.7915 & 0.13467650 & 0.70597(29) & 0.65418(67) &  0.9265(10)(2)    & 0.9298(15)       & -0.250(84) & 0.63 \\
          & 12  &  7.0688 & 0.13420891 & 0.69553(40) & 0.64543(63) &  0.9276(11)(4)    &                  &            &      \\  \midrule
  \multirow{4}{*}{2.0120} 
         & 6   &  6.2735  & 0.13557130 & 0.69013(32) & 0.62979(37) &  0.9139(7)(13)    &  \multirow{4}{*}{0.9149(10)}  & \multirow{4}{*}{$-0.006(69)$} & \multirow{4}{*}{0.59} \\
         & 8   &  6.4680  & 0.13523620 & 0.68107(28) & 0.62341(43) &  0.91518(74)(17)  &                  &            &      \\
         & 12  &  6.72995 & 0.13475973 & 0.67113(43) & 0.61452(49) &  0.91523(93)(41)  &                  &            &      \\
         & 16  &  6.93460 & 0.13441209 & 0.66627(31) & 0.60924(66) &  0.9141(11)(3)    &                  &            &      \\
  \bottomrule
  \end{tabular}
  \caption{Results for $\ZP$, $\SigmaPI$, and $\sigmaP$ in the $\HEs$ scheme.
           The last three columns quote the value of $\sigmaP$ obtained from a
           continuum-limit extrapolation fitting all three points linearly in
           $(a/L)^2$, the corresponding slope parameters, and the values of $\chi^2$
           per degree of freedom (note that the numbers of degrees of freedom is always~1,
           save for the case $u=2.0120$ where it is~2).
          }
  \label{tab:ZP_SF}
\end{table}
\end{landscape}

\cleardoublepage
\begin{landscape}
\begin{table}
  \vskip-1em
  \centering
  \small
  \begin{tabular}{ll>{\ttfamily}llllllllllll}\toprule
          Fit & $\left(\frac{L}{a}\right)_{\rm\scriptscriptstyle min}$ & type & $n_s$ & $n_\rho$ & $\chi^2/\text{dof}$ & $k=0$ & $k=1$ & $k=2$ & $k=3$ & $k=4$ & $k=5$ \\ \midrule
\multicolumn{3}{l}{$u_k$}   & -- & -- & -- & 2.0120 & 1.7126(31) & 1.4939(38) & 1.3264(38) & 1.1936(35) & 1.0856(32) \\ \midrule
$\sigmaP$:\emph{u-by-u} & 6 & FITA  & $4$ &  -- & 4.7/5   & 0.91522(98) & 0.8528(13) & 0.8034(16) & 0.7626(17) & 0.7280(17) & 0.6980(18) \\
                  &   & FITA  & $5$ &  -- & 4.4/4   & 0.91500(99) & 0.8533(13) & 0.8036(15) & 0.7624(17) & 0.7278(18) & 0.6985(19) \\
                  &   & FITB  & $4$ &  -- & 5.1/6   & 0.91540(99) & 0.8526(13) & 0.8030(14) & 0.7622(16) & 0.7278(17) & 0.6982(19) \\
                  &   & FITB  & $5$ &  -- & 3.3/5   & 0.91515(97) & 0.8529(12) & 0.8036(15) & 0.7627(16) & 0.7280(17) & 0.6981(18) \\ \midrule
$\sigmaP$:\emph{global} & 6 & FITA  & $4$ & $2$ & 19.1/21 & 0.91682(63) & 0.85406(93)& 0.8040(12) & 0.7625(14) & 0.7274(15) & 0.6972(16) \\
                  &   & FITA  & $5$ & $3$ & 13.7/19 & 0.91550(85) & 0.8528(12) & 0.8031(14) & 0.7623(15) & 0.7278(17) & 0.6982(19) \\
                  &   & FITB  & $4$ & $2$ & 20.2/22 & 0.91698(72) & 0.8540(11) & 0.8037(13) & 0.7623(15) & 0.7273(16) & 0.6972(17) \\
                  &   & FITB  & $5$ & $3$ & 13.8/20 & 0.91551(94) & 0.8527(13) & 0.8031(15) & 0.7623(17) & 0.7278(18) & 0.6981(19) \\ \cmidrule(lr){2-12}
                  & 8 & FITA  & $4$ & $2$ & 12.9/13 & 0.9157(10)  & 0.8525(16) & 0.8021(20) & 0.7604(23) & 0.7252(24) & 0.6948(23) \\
                  &   & FITA  & $5$ & $3$ & 7.7/11  & 0.9142(11)  & 0.8514(15) & 0.8017(17) & 0.7608(20) & 0.7264(22) & 0.6970(25) \\ 
                  &   & FITB  & $4$ & $2$ & 14.3/14 & 0.91595(91) & 0.8524(15) & 0.8018(18) & 0.7602(21) & 0.7251(23) & 0.6950(24) \\ 
                  &   & FITB  & $5$ & $3$ & 8.4/12  & 0.9143(12)  & 0.8513(16) & 0.8016(19) & 0.7608(21) & 0.7264(24) & 0.6967(26) \\ \midrule
                      $\tau$:\emph{u-by-u} & 6 & FITB  & $2$ &  -- & 6.2/7   & 0.91794(86) & 0.8552(13) & 0.8051(15) & 0.7636(16) & 0.7286(17) & 0.6984(17) \\
                  &   & FITB  & $3$ &  -- & 3.5/6   & 0.9159(11)  & 0.8533(14) & 0.8036(16) & 0.7628(17) & 0.7282(18) & 0.6985(19) \\
                  &   & FITB  & $4$ &  -- & 3.3/5   & 0.9156(11)  & 0.8532(13) & 0.8037(16) & 0.7629(17) & 0.7282(17) & 0.6983(18) \\ 
                  &   & FITB  & $5$ &  -- & 2.4/4   & 0.9151(10)  & 0.8534(13) & 0.8036(15) & 0.7624(16) & 0.7279(16) & 0.6984(17) \\ \midrule
    $\tau$:\emph{global} & 6 & FITA  & $2$ & $2$ & 18.5/22 & 0.91764(88) & 0.8547(13) & 0.8045(15) & 0.7630(16) & 0.7279(17) & 0.6977(18) \\ 
                  &   & FITB  & $2$ & $2$ & 18.5/23 & 0.91763(98) & 0.8547(14) & 0.8044(16) & 0.7630(17) & 0.7279(18) & 0.6971(18) \\
                  &   & FITB  & $3$ & $2$ & 18.4/22 & 0.91751(86) & 0.8546(12) & 0.8044(15) & 0.7629(16) & 0.7278(17) & 0.6977(18) \\
                  &   & FITB  & $2$ & $3$ & 16.1/22 & 0.91775(96) & 0.8550(13) & 0.8048(15) & 0.7631(17) & 0.7283(17) & 0.6981(18) \\
                  &   & FITB  & $3$ & $3$ & 13.0/21 & 0.9155(12)  & 0.8527(14) & 0.8031(15) & 0.7623(16) & 0.7278(17) & 0.6981(18) \\ \cmidrule(lr){2-12}
                  & 8 & FITA  & $2$ & $2$ & 12.0/14 & 0.9166(11)  & 0.8532(16) & 0.8026(19) & 0.7608(21) & 0.7256(22) & 0.6954(24) \\
                  &   & FITB* & $2$ & $2$ & 12.0/15 & 0.9165(12)  & 0.8530(17) & 0.8025(20) & 0.7608(21) & 0.7257(22) & 0.6955(23) \\
                  &   & FITB  & $3$ & $2$ & 12.0/14 & 0.9166(10)  & 0.8531(15) & 0.8025(18) & 0.7608(21) & 0.7257(22) & 0.6954(23) \\
                  &   & FITB  & $2$ & $3$ & 10.0/14 & 0.9165(12)  & 0.8532(16) & 0.8027(19) & 0.7611(21) & 0.7260(21) & 0.6958(22) \\
                  &   & FITB  & $3$ & $3$ & 7.3/13  & 0.9144(14)  & 0.8514(17) & 0.8017(19) & 0.7609(21) & 0.7264(23) & 0.6968(24) \\ \midrule
  PT prediction   & --& s1,s2 & --  & --  & 13/8    & 0.90797     & 0.84057    & 0.78808    & 0.74552    & 0.70998    & 0.67966    \\
    \bottomrule
  \end{tabular}
  \caption{Mass-ratios $R^{(k)}$, cf. Eqs.~(\ref{eq:Rk-sigma}) and (\ref{eq:recursion2}), as obtained from
           different analysis procedures. The quoted values of $u_k$ have
           been obtained using Eq.~\eqref{eq:betasf} with $b^{\rm eff}_3$.
          }
  \label{tab:mrat}
\end{table}
\end{landscape}

\cleardoublepage
\begin{table}
\centering
\small
\begin{tabular}{C@{\hspace{5mm}}CCC@{\hspace{5mm}}CLL}
\toprule
    \uGF   & L/a& \beta    & \kappa     & \ZP(g_0^2,L/a)& \ZP(g_0^2,2L/a) & \SigmaP(g_0^2,L/a) \\\midrule
2.1269(15) & 8  & 5.371500 & 0.13362120 & 0.73275(26)   &  0.67666(55)    & 0.92345(82)  \\
2.1229(12) & 12 & 5.543070 & 0.13331407 & 0.71301(33)   &  0.65748(85)    & 0.9221(12)   \\
2.1257(25) & 16 & 5.700000 & 0.13304840 & 0.70248(31)   &  0.64369(85)    & 0.9163(13)   \\\midrule
2.3913(15) & 8  & 5.071000 & 0.13421678 & 0.71024(32)   &  0.64872(59)    & 0.91338(93)  \\
2.3912(10) & 12 & 5.242465 & 0.13387635 & 0.69064(30)   &  0.62808(99)    & 0.9094(15)   \\
2.3900(32) & 16 & 5.400000 & 0.13357851 & 0.67890(35)   &  0.61636(95)    & 0.9079(15)   \\\midrule                                                                          
2.7365(14) & 8  & 4.764900 & 0.13488555 & 0.68196(32)   &  0.61339(69)    & 0.8995(11)   \\
2.7390(14) & 12 & 4.938726 & 0.13450761 & 0.66137(42)   &  0.59091(84)    & 0.8935(14)   \\
2.7359(36) & 16 & 5.100000 & 0.13416889 & 0.65087(41)   &  0.58199(92)    & 0.8942(15)   \\\midrule                                                                          
3.2022(16) & 8  & 4.457600 & 0.13560675 & 0.64779(34)   &  0.56891(78)    & 0.8782(13)   \\ 
3.2053(17) & 12 & 4.634654 & 0.13519986 & 0.62622(39)   &  0.54748(98)    & 0.8743(16)   \\
3.2029(49) & 16 & 4.800000 & 0.13482139 & 0.61735(43)   &  0.5382(11)     & 0.8717(19)   \\\midrule                                                                          
3.8620(20) & 8  & 4.151900 & 0.13632589 & 0.60377(37)   &  0.51002(78)    & 0.8447(14)   \\
3.8635(21) & 12 & 4.331660 & 0.13592664 & 0.58285(49)   &  0.4887(14)     & 0.8384(26)   \\
3.8643(64) & 16 & 4.500000 & 0.13552582 & 0.57420(48)   &  0.4822(14)     & 0.8398(25)   \\\midrule                                                                          
4.4855(25) & 8  & 3.947900 & 0.13674684 & 0.56568(39)   &  0.4570(11)     & 0.8079(21)   \\
4.4867(28) & 12 & 4.128217 & 0.13640300 & 0.54608(62)   &  0.4331(10)     & 0.7931(21)   \\ 
4.4901(75) & 16 & 4.300000 & 0.13600821 & 0.54030(55)   &  0.4266(16)     & 0.7895(30)   \\\midrule                                                                          
5.2928(28) & 8  & 3.754890 & 0.13701929 & 0.52174(43)   &  0.3928(29)     & 0.7528(57)   \\
5.2972(36) & 12 & 3.936816 & 0.13679805 & 0.50367(57)   &  0.3642(21)     & 0.7231(41)   \\ 
5.301(14)  & 16 & 4.100000 & 0.13647301 & 0.49847(79)   &  0.3579(28)     & 0.7179(54)   \\
\bottomrule
\end{tabular}
\caption{Results for $\ZP$ and $\SigmaP$ in the $\LEs$ scheme.
        }
\label{tab:ZP_GF}
\end{table}

\begin{table}[p]
  \vskip-3em
  \centering
  \small\footnotesize
  \begin{tabular}{CCCCCCCLLCCC}\toprule
   \uSF   & L/a & \beta    & \kappa     & s & \Nms & \dfrac{\tms}{\tau} 
                                                        & \dfrac{\tint[\ZP]}{\tms} 
                                                                   & 10^{2} Lm_1 
                                                                              & f_1 \\\midrule
  1.11000 & 6   &  8.5403  & 0.13233610 & 1 & 5000 & 5  & 0.50(2)  & -0.74(4) & 0.9706(11)  \\
          &     &          &            & 2 & 5000 & 5  & 0.76(7)  & -1.60(3) & 0.8929(12)  \\
          & 8   &  8.7325  & 0.13213380 & 1 & 5000 & 5  & 0.58(5)  & -0.97(3) & 0.9526(10)  \\
          &     &          &            & 2 & 2683 & 5  & 0.92(13) & -1.81(3) & 0.8795(21)  \\
          & 12  &  8.9950  & 0.13186210 & 1 & 2769 & 5  & 0.72(9)  & -0.64(3) & 0.9161(16)  \\
          &     &          &            & 2 & 1576 & 5  & 1.71(39) & -1.14(3) & 0.8412(40)  \\  \midrule
  1.18446 & 6   &  8.2170  & 0.13269030 & 1 & 5000 & 5  & 0.56(4)  & -0.74(4) & 0.9428(11)  \\
          &     &          &            & 2 & 5000 & 5  & 0.75(7)  & -1.63(3) & 0.8690(12)  \\
          & 8   &  8.4044  & 0.13247670 & 1 & 5000 & 5  & 0.49(1)  & -1.00(3) & 0.9276(10)  \\
          &     &          &            & 2 & 2314 & 5  & 0.89(13) & -1.91(4) & 0.8525(29)  \\
          & 12  &  8.6769  & 0.13217153 & 1 & 2476 & 5  & 0.60(7)  & -0.76(4) & 0.8938(17)  \\
          &     &          &            & 2 & 3759 & 5  & 2.45(46) & -1.25(2) & 0.8176(23)  \\  \midrule
  1.26569 & 6   &  7.9091  & 0.13305720 & 1 & 5000 & 5  & 0.50(1)  & -0.74(4) & 0.9160(11)  \\
          &     &          &            & 2 & 5000 & 5  & 0.72(7)  & -1.73(3) & 0.8393(13)  \\
          & 8   &  8.0929  & 0.13283120 & 1 & 5000 & 5  & 0.54(4)  & -1.08(4) & 0.9005(11)  \\ 
          &     &          &            & 2 & 2273 & 5  & 1.01(16) & -1.98(4) & 0.8252(21)  \\
          & 12  &  8.3730  & 0.13249231 & 1 & 2729 & 5  & 0.71(9)  & -0.76(4) & 0.8654(18)  \\
          &     &          &            & 2 & 3749 & 5  & 2.20(40) & -1.35(2) & 0.7926(32)  \\  \midrule
  1.3627  & 6   &  7.5909  & 0.13346930 & 1 & 5000 & 5  & 0.50(2)  & -0.81(5) & 0.8877(11)  \\
          &     &          &            & 2 & 5000 & 5  & 0.69(6)  & -1.63(3) & 0.8034(15)  \\
          & 8   &  7.7723  & 0.13322830 & 1 & 5000 & 5  & 0.57(5)  & -1.18(4) & 0.8725(11)  \\
          &     &          &            & 2 & 2163 & 5  & 0.86(13) & -2.12(5) & 0.7892(36)  \\
          & 12  &  8.0578  & 0.13285365 & 1 & 2448 & 5  & 0.71(9)  & -0.81(4) & 0.8423(17)  \\
          &     &          &            & 2 & 3762 & 5  & 2.42(45) & -1.49(2) & 0.7594(29)  \\    \bottomrule
  \end{tabular}
  \caption{Details for SF simulations of $(s\times L/a)^4$ lattices with
           vanishing background field and plaquette gauge action, cf.
           \res{sec:sf}.  Each trajectory has a length of $\tau=2$ MD
           units, and lines of constant physics (fixed $\uSF$) are set with
           background field as reported in Refs.~\cite{Brida:2016flw,SFpaper}
           (continues on the next page).
           }
  \label{tab:SFruns}
\end{table}
\addtocounter{table}{-1}
\begin{table}[p]
  \vskip-3em
  \centering
  \small\footnotesize
  \begin{tabular}{CCCCCCCLLCCC}\toprule
   \uSF   & L/a & \beta    & \kappa     & s & \Nms & \dfrac{\tms}{\tau} 
                                                        & \dfrac{\tint[\ZP]}{\tms} 
                                                                   & 10^{2} Lm_1 
                                                                              & f_1 \\\midrule
  1.4808  & 6   &  7.2618  & 0.13393370 & 1 & 5000 & 5  & 0.49(1)  & -0.87(5) & 0.8547(11)  \\ 
          &     &          &            & 2 & 5000 & 5  & 0.87(9)  & -1.50(4) & 0.7658(16)  \\
          & 8   &  7.4424  & 0.13367450 & 1 & 5000 & 5  & 0.60(5)  & -1.15(4) & 0.8385(11)  \\
          &     &          &            & 2 & 4500 & 5  & 0.99(12) & -2.09(3) & 0.7529(17)  \\
          & 12  &  7.7299  & 0.13326353 & 1 & 2710 & 5  & 0.67(8)  & -0.88(4) & 0.8119(20)  \\
          &     &          &            & 2 & 6343 & 5  & 2.33(34) & -1.49(3) & 0.7226(23)  \\  \midrule
  1.6173  & 6   &  6.9433  & 0.13442200 & 1 & 5000 & 5  & 0.50(3)  & -0.76(5) & 0.8152(13)  \\  
          &     &          &            & 2 & 5000 & 5  & 0.68(6)  & -1.39(4) & 0.7219(15)  \\
          & 8   &  7.1254  & 0.13414180 & 1 & 5000 & 5  & 0.56(4)  & -1.12(5) & 0.8058(12)  \\ 
          &     &          &            & 2 & 2041 & 5  & 0.87(14) & -2.06(5) & 0.7172(29)  \\
          & 12  &  7.4107  & 0.13369922 & 1 & 2535 & 5  & 0.80(11) & -0.87(5) & 0.7740(20)  \\
          &     &          &            & 2 & 3412 & 5  & 3.88(91) & -1.59(3) & 0.6851(31)  \\  \midrule
  1.7943  & 6   &  6.6050  & 0.13498290 & 1 & 5000 & 5  & 0.49(1)  & -0.80(6) & 0.7743(13)  \\  
          &     &          &            & 2 & 5000 & 5  & 0.76(7)  & -1.18(4) & 0.6742(17)  \\
          & 8   &  6.7915  & 0.13467650 & 1 & 5000 & 5  & 0.57(5)  & -1.17(6) & 0.7637(13)  \\
          &     &          &            & 2 & 1807 & 5  & 1.35(26) & -2.10(6) & 0.6688(40)  \\
          & 12  &  7.0688  & 0.13420891 & 1 & 2339 & 5  & 0.68(8)  & -0.83(6) & 0.7311(22)  \\
          &     &          &            & 2 & 2607 & 5  & 1.94(38) & -1.48(4) & 0.6351(33)  \\  \midrule
  2.0120  & 6   &  6.2735  & 0.13557130 & 1 & 5000 & 5  & 0.53(4)  & -0.69(7) & 0.7233(13)  \\
          &     &          &            & 2 & 4435 & 5  & 0.85(9)  & -0.92(5) & 0.6209(21)  \\
          & 8   &  6.4680  & 0.13523620 & 1 & 5000 & 5  & 0.53(4)  & -1.14(6) & 0.7204(14)  \\
          &     &          &            & 2 & 4048 & 5  & 1.04(13) & -1.93(5) & 0.6155(20)  \\
          & 10  &  6.60959 & 0.13497763 & 1 & 6240 & 5  & 0.72(6)  & -1.08(4) & 0.7018(14)  \\
          & 12  &  6.72995 & 0.13475973 & 1 & 3000 & 5  & 0.86(11) & -0.80(6) & 0.6915(19)  \\
          &     &          &            & 2 & 5094 & 5  & 2.42(40) & -1.44(3) & 0.5808(55)  \\
          &     &          &            & 2 & 512  & 10 & 1.14(34) & -1.33(10)& 0.5833(25)  \\
          & 16  &  6.93460 & 0.13441209 & 1 & 4604 & 10 & 0.73(7)  & -0.51(4) & 0.6655(16)  \\
          &     &          &            & 2 & 2325 & 10 & 1.69(3)  & -1.01(4) & 0.5583(44)  \\
  \bottomrule
  \end{tabular}
  \caption{(continued)}
\end{table}

\begin{table}[p]
  \vskip-5em
  \centering
  \small\footnotesize
  \begin{tabular}{CCCCCCRLLCC}\toprule
     \uGF & L/a & \beta     & \kappa       & s & \Nms & \dfrac{\tms}{\text{MD}} 
                                                           & \dfrac{\tint[\ZP]}{\tms} 
                                                                       & 10^{2} Lm_1 
                                                                                  & f_1 \\\midrule
  2.1257  &  8  &  5.371500 &  0.13362120  & 1 & 5001 & 10 &  0.53(4)  & +0.03(5) & 0.8471(14)   \\  
          &     &           &              & 2 & 2001 & 10 &  1.21(22) & +0.52(5) & 0.7175(40)   \\  
          &  12 &  5.543070 &  0.13331407  & 1 & 8000 & 5  &  1.35(14) & +0.13(3) & 0.7962(15)   \\  
          &     &           &              & 2 & 2400 & 10 &  3.17(58) & +0.62(4) & 0.6618(53)   \\  
          &  16 &  5.700000 &  0.13304840  & 1 & 7001 & 10 &  1.26(14) & -0.04(3) & 0.7697(19)   \\  
          &     &           &              & 2 & 500  & 100&  0.54(11) & +0.38(7) & 0.6414(51)   \\\midrule  
  2.3900  &  8  &  5.071000 &  0.13421678  & 1 & 5001 & 10 &  0.58(5)  & +0.02(5) & 0.8092(15)   \\  
          &     &           &              & 2 & 2001 & 10 &  1.09(19) & +0.66(6) & 0.6677(37)   \\  
          &  12 &  5.242465 &  0.13387635  & 1 & 8000 & 5  &  1.22(12) & +0.05(4) & 0.7602(18)   \\  
          &     &           &              & 2 & 2400 & 10 &  3.21(59) & +0.64(4) & 0.6145(53)   \\  
          &  16 &  5.400000 &  0.13357851  & 1 & 6001 & 10 &  1.16(13) & +0.07(3) & 0.7263(23)   \\  
          &     &           &              & 2 & 500  & 100&  0.68(13) & +0.59(9) & 0.5970(42)   \\\midrule  
  2.7359  &  8  &  4.764900 &  0.13488555  & 1 & 5001 & 10 &  0.59(5)  & +0.04(6) & 0.7675(14)   \\  
          &     &           &              & 2 & 2001 & 10 &  1.36(26) & +0.91(7) & 0.6185(36)   \\  
          &  12 &  4.938726 &  0.13450761  & 1 & 5001 & 5  &  1.13(13) & +0.06(6) & 0.7123(25)   \\  
          &     &           &              & 2 & 2400 & 10 &  2.45(41) & +0.78(5) & 0.5607(57)   \\  
          &  16 &  5.100000 &  0.13416889  & 1 & 6001 & 10 &  1.23(14) & +0.04(3) & 0.6814(21)   \\  
          &     &           &              & 2 & 500  & 100&  0.74(14) & +0.67(8) & 0.5385(42)   \\\midrule  
  3.2029  &  8  &  4.457600 &  0.13560675  & 1 & 5001 & 10 &  0.55(4)  & +0.09(7) & 0.7239(16)   \\  
          &     &           &              & 2 & 2001 & 10 &  1.20(22) & +1.10(9) & 0.5524(65)   \\  
          &  12 &  4.634654 &  0.13519986  & 1 & 5001 & 5  &  1.05(12) & +0.06(6) & 0.6684(30)   \\  
          &     &           &              & 2 & 2400 & 10 &  2.88(51) & +0.68(6) & 0.5052(57)   \\  
          &  16 &  4.800000 &  0.13482139  & 1 & 5000 & 10 &  1.31(17) & +0.12(4) & 0.6307(25)   \\  
          &     &           &              & 2 & 2000 & 20 &  2.47(44) & +0.79(5) & 0.4813(51)   \\\midrule  
  3.8643  &  8  &  4.151900 &  0.13632589  & 1 & 5001 & 10 &  0.58(5)  & -0.07(9) & 0.6798(19)   \\  
          &     &           &              & 2 & 2001 & 10 &  1.23(23) & +1.58(10)& 0.4866(51)   \\  
          &  12 &  4.331660 &  0.13592664  & 1 & 5001 & 5  &  1.19(15) & -0.10(8) & 0.6190(29)   \\  
          &     &           &              & 2 & 2400 & 10 &  3.17(58) & +0.77(7) & 0.4381(78)   \\  
          &  16 &  4.500000 &  0.13552582  & 1 & 5000 & 10 &  1.30(17) & +0.00(5) & 0.5805(24)   \\
          &     &           &              & 2 & 2000 & 20 &  3.07(60) & +0.68(6) & 0.3921(80)   \\  \bottomrule
  \end{tabular}
  \caption{Details for SF simulations of $(s\times L/a)^4$ lattices with
           vanishing background field and tree-level Symanzik improved gauge
           action, cf. \res{sec:gf}.  Lines of constant physics (fixed
           $\uGF$) are set as reported in Ref.~\cite{DallaBrida:2016kgh}
           (continues on the next page).
          }
  \label{tab:GFruns}
\end{table}
\addtocounter{table}{-1}
\begin{table}[p]
  \vskip-5em
  \centering
  \small\footnotesize
  \begin{tabular}{CCCCCCRLLCC}\toprule
     \uGF & L/a & \beta     & \kappa       & s & \Nms & \dfrac{\tms}{\text{MD}} 
                                                           & \dfrac{\tint[\ZP]}{\tms} 
                                                                       & 10^{2} Lm_1 
                                                                                  & f_1 \\\midrule
  4.4901  &  8  &  3.947900 &  0.13674684  & 1 & 5001 & 10 &  0.61(6)  & -0.11(10)& 0.6385(19)   \\  
          &     &           &              & 2 & 2001 & 10 &  2.04(46) & +2.02(11)& 0.4297(56)   \\  
          &  12 &  4.128217 &  0.13640300  & 1 & 5001 & 5  &  1.84(27) & -0.07(9) & 0.5802(33)   \\  
          &     &           &              & 2 & 4127 & 10 &  3.81(60) & +1.03(7) & 0.3622(55)   \\  
          &  16 &  4.300000 &  0.13600821  & 1 & 5000 & 10 &  1.56(22) & -0.00(6) & 0.5343(27)   \\  
          &     &           &              & 2 & 3439 & 20 &  5.1(1.0) & +0.86(5) & 0.3338(57)   \\\midrule  
  5.3010  &  8  &  3.754890 &  0.13701929  & 1 & 5001 & 5  &  0.84(9)  & +0.23(14)& 0.6094(30)   \\  
          &     &           &              & 2 & 2001 & 10 &  8.6(3.4) & +3.17(16)& 0.3474(95)   \\  
          &  12 &  3.936816 &  0.13679805  & 1 & 5001 & 5  &  1.13(14) & -0.18(10)& 0.5326(33)   \\  
          &     &           &              & 2 & 3682 & 10 &  7.6(2.2) & +1.54(8) & 0.2878(78)   \\  
          &  16 &  4.100000 &  0.13647301  & 1 & 3200 & 10 &  1.48(24) & -0.04(8) & 0.4898(43)   \\  
          &     &           &              & 2 & 2674 & 20 & 11.8(4.7) & +1.04(7) & 0.2612(72)   \\\midrule    
  5.8673  &  8  &  3.653850 &  0.13707221  & 1 & 5001 & 5  &  8.97(10) & +0.10(16)& 0.5941(31)   \\  
          &     &           &              & 2 & 2000 & 10 &  3.7(1.0) & +4.82(22)& 0.3168(72)   \\  
          &  12 &  3.833254 &  0.13696774  & 1 & 5001 & 5  &  2.00(31) & +0.14(11)& 0.5139(37)   \\  
          &     &           &              & 2 & 2400 & 10 & 11.8(4.8) & +2.03(13)& 0.2360(79)   \\  
          &  16 &  4.000000 &  0.13668396  & 1 & 4602 & 10 &  1.45(20) & +0.06(8) & 0.4663(35)   \\  
          &     &           &              & 2 & 1404 & 20 & 32(20)    & +1.71(12)& 0.1895(16)   \\\midrule    
  6.5489  &  8  &  3.556470 &  0.13703245  & 1 & 5001 & 5  &  1.03(11) & -0.18(19)& 0.5871(36)   \\  
          &     &           &              & 2 & 2000 & 10 &  4.7(1.5) & +7.69(34)& 0.2304(72)   \\  
          &  12 &  3.735394 &  0.13708263  & 1 & 5001 & 5  &  1.53(21) & -0.06(14)& 0.4905(43)   \\  
          &     &           &              & 2 & 3000 & 10 & 19.2(8.7) & +2.74(16)& 0.1864(69)   \\  
          &  16 &  3.900000 &  0.13687202  & 1 & 4600 & 10 &  1.60(23) & +0.08(8) & 0.4360(35)   \\  
          &     &           &              & 2 & 1205 & 20 & 11.4(5.9) & +1.87(21)& 0.1656(91)   \\  
  \bottomrule
  \end{tabular}
  \caption{(continued)}
\end{table}

\end{appendix}

\cleardoublepage
\bibliographystyle{JHEPjus}
\providecommand{\href}[2]{#2}

\end{document}